\documentclass[twocolumn,showpacs,superscriptaddress,preprintnumbers,amsmath,amssymb,epsfig,color]{revtex4}

\usepackage{color}
\usepackage{graphicx}
\usepackage{dcolumn}
\usepackage{bm}

\newcommand{\beq}{\begin{equation}}
\newcommand{\eeq}{\end{equation}}
\newcommand{\bea}{\begin{eqnarray}}
\newcommand{\ena}{\end{eqnarray}}
\newcommand{\etal}{{\it et al.}}

\newcommand{\lsim}{\mathrel{\mathop{\kern 0pt \rlap
{\raise.2ex\hbox{$<$}}}
\lower.9ex\hbox{\kern-.190em $\sim$}}}
\newcommand{\gsim}{\mathrel{\mathop{\kern 0pt \rlap
{\raise.2ex\hbox{$>$}}}
\lower.9ex\hbox{\kern-.190em $\sim$}}}

\newcommand{\sigmav}{\langle \sigma_{\rm ann} v \rangle_0}
\newcommand{\sigmavint}{\langle \sigma_{\rm ann} v \rangle_{\rm int}}
\newcommand{\relic}{\Omega_\chi h^2}

\newcommand{\app}[3]{Astropart.\ Phys.\ {\bf #1}, #3 (#2)}

\newcommand{\hepph}[1]{{\tt hep-ph/#1}}
\newcommand{\astroph}[1]{{\tt astro-ph/#1}}

\newcommand{\plb}[3]{Phys.\ Lett.\ B\ {\bf #1}, #3 (#2)}
\newcommand{\npb}[3]{Nucl.\ Phys.\ B\ {\bf #1}, #3 (#2)}
\newcommand{\npps}[3]{Nucl.\ Phys.\ Proc.\ Suppl.\ {\bf #1}, #3 (#2)}

\renewcommand{\apj}[3]{Astrophys.\ J.\ {\bf #1}, #3 (#2)}
\newcommand{\aeta}[3]{Astron.\  Astrophys.\ {\bf #1}, #3 (#2)}

\renewcommand{\prd}[3]{Phys.\ Rev.\ D\ {\bf #1}, #3 (#2)}

\begin{document}

\preprint{DFTT-32-02,KIAS-P02069}

\title{Antiprotons in cosmic rays from neutralino annihilation}

\author{F. Donato}
\affiliation{Dipartimento di Fisica Teorica, Universit\`a di Torino \\
Istituto Nazionale di Fisica Nucleare, Sezione di Torino \\ via
P. Giuria 1, I--10125 Torino, Italy} 
\email{donato,fornengo@to.infn.it}

\author{N. Fornengo}
\affiliation{Dipartimento di Fisica Teorica, Universit\`a di Torino \\
Istituto Nazionale di Fisica Nucleare, Sezione di Torino \\ via
P. Giuria 1, I--10125 Torino, Italy} 
\affiliation{School of Physics,
Korea Institute for Advanced Study\\ 207-43 Cheongryangri-dong,
Dongdaemun-gu, Seoul 130-012, Korea}

\author{D. Maurin} 
\affiliation{Service d'Astrophysique,
SAp CEA-Saclay, F-91191 Gif-sur-Yvette CEDEX, France}
\email{dmaurin@cea.fr}

\author{P. Salati} 
\affiliation{Laboratoire de
Physique Th\'eorique {\sc lapth}, Annecy--le--Vieux, 74941, and
Universit\'e de Savoie, Chamb\'ery, 73011, France}
\email{salati,taillet@lapp.in2p3.fr}

\author{R. Taillet} 
\affiliation{Laboratoire de
Physique Th\'eorique {\sc lapth}, Annecy--le--Vieux, 74941, and
Universit\'e de Savoie, Chamb\'ery, 73011, France}
\email{salati,taillet@lapp.in2p3.fr}

\date{\today}

\begin{abstract}
We calculate the antiproton flux due to relic neutralino
annihilations, in a two--dimensional diffusion model compatible with
stable and radioactive cosmic ray nuclei.
We find that the uncertainty in the primary flux induced by the
propagation parameters alone is about two orders of magnitude at low
energies, and it is mainly determined by the lack of knowledge on the
thickness of the diffusive halo.  On the contrary, different dark
matter density profiles do not significantly alter the flux: a NFW
distribution produces fluxes which are at most 20\% higher than an
isothermal sphere. The most conservative choice for propagation
parameters and dark matter distribution normalization, together with
current data on cosmic antiprotons, cannot lead to any definitive
constraint on the supersymmetric parameter space, neither in a
low--energy effective MSSM, or in a minimal SUGRA scheme. However, if
the best choice for propagation parameters -- corresponding to a
diffusive halo of $L=4$ kpc -- is adopted, some supersymmetric
configurations with the neutralino mass $m_\chi \lsim 100$ GeV should
be considered as excluded. An enhancement flux factor - due for
instance to a clumpy dark halo or to a higher local dark matter
density - would imply a more severe cut on the supersymmetric
parameters.
\end{abstract}

\pacs{95.35.+d,98.35.Gi,98.35.Pr,96.40.-z,98.70.Sa,11.30.Pb,12.60.Jv,95.30.Cq}

\maketitle


\section{Introduction}
\label{sec:introduction}

The recent {\sc wmap} measurements of the Cosmic Microwave Background (CMB)
anisotropies \cite{cmb} point towards a flat universe with a fraction
$\Omega_{\Lambda} \simeq 0.7$ of the closure density in the form of a
negative pressure component -- such as a cosmological constant or a
scalar field -- while the remaining $\Omega_{\rm m} \simeq 0.3$ is
matter. These conclusions are independently reached from the
determination of the relation between the luminosity distance and the
redshift of supernovae SNeIa \cite{sn} on the one hand and from the
large scale structure (LSS) information from Galaxy and cluster
surveys \cite{lss}. The {\sc wmap} values of $\Omega_{\rm m} = 0.27 \pm
0.04$ and $\Omega_{\rm B} = 0.044 \pm 0.004$ indicate that most of the
matter is non--baryonic. The amount of baryonic matter $\Omega_{\rm
B}$ deduced from CMB is in perfect agreement with the results from
primordial nucleosynthesis and observations of the deuterium abundance
in quasar absorption lines \cite{nucleosynthesis}.

The nature of this astronomical dark matter has been challenging
physicists for several decades and is still unresolved. The favoured
candidate is a weakly--interacting massive particle (WIMP). The
so--called neutralino naturally arises in the framework of supersymmetric
theories as the lightest combination of neutral higgsinos and gauginos.
Large efforts have been devoted to pin down these evading species
\cite{SUSY_DM_general}. Experimental techniques
\cite{direct1,direct0,direct2} have been devised in order to be
sensitive to the recoil energy which a neutralino may deposit as it
crosses a terrestrial detector.
The annihilation photons from the neutralinos that populate the
Milky--Way halo \cite{gamma_neutralino_MW} or extra--Galactic
systems \cite{gamma_neutralino_M87} are under scrutiny. As a matter
of fact, a gamma--ray excess has been recently reported by {\sc hegra} in the
direction of the giant elliptical M87 \cite{excess_M87}. Antimatter
cosmic--ray particles are also expected from neutralino annihilations
inside our Galaxy. A subtle feature in the positron spectrum has actually
been measured by the {\sc heat} 
Collaboration \cite{heat} for energies beyond 7 GeV.

This work is devoted to cosmic--ray antiprotons whose energy spectrum
has already been measured with some accuracy. Much larger statistics
will soon be collected by the {\sc ams} collaboration on board the {\sc iss} 
by the {\sc bess}-Polar long duration, balloon experiment and by the 
{\sc pamela} 
satellite.
Secondary antiprotons are naturally produced by the spallation
of primary nuclei -- mostly cosmic--ray protons and helions -- on
the diffuse gas of the Milky--Way ridge. If neutralinos pervade our
Galaxy, a primary component adds up to that secondary distribution.
The spectral distortion that ensues is expected a priori in the
low--energy region for mere kinematic reasons \cite{Bottino_Salati}~:
unlike for a neutralino annihilation, the center--of--mass frame of
a spallation event is not at rest with respect to the Galaxy.
In principle, an excess of low--energy antiprotons is the signature
of an unconventional production -- either neutralino annihilation
or small black holes evaporation \cite{pbar_pbh} for instance.
However, because antiprotons undergo inelastic yet non--annihilating
collisions with the interstellar material, the high--energy particles 
tend to lose energy and to populate the low--energy tail of the spectrum
that consequently is much flatter \cite{bergstrom} than previously
estimated. 
This motivated the search of other cosmic--ray signatures
such as antideuterons \cite{dbar,dbar_pbh}.
Antiproton production from primary cosmic--ray spallations is the
natural background to any unconventional excess that would signal
for instance the presence of the putative neutralinos. The detailed
calculation of that secondary component \cite{PaperII} has required the
determination of the propagation--diffusion parameters that are
consistent with the B/C data \cite{PaperI}. By varying
those parameters over the entire range allowed by the cosmic--ray
nuclei measurements, the theoretical uncertainty on the antiproton
secondary flux has been found to be 9\% from 100 MeV to 1 GeV. It
reaches a maximum of 24\% at 10 GeV and decreases to 10\% at 100 GeV.
This small scatter in the secondary antiproton spectrum is not
surprising. Cosmic--ray nuclei such as LiBeB and secondary antiprotons
are both manufactured in the same place -- the interstellar gas of
the Galactic disk -- through the same production mechanism -- the
spallation of primaries.

The aim of this article is to calculate  the supersymmetric cosmic--ray 
antiproton flux that arise from the diffusion--propagation parameter space
and to estimate the uncertainties due to its spread.  Since neutralinos
annihilate all over the Milky Way and are not confined to the disk
alone, we anticipate that the uncertainty in that primary component
will be much larger than for secondaries.

The discusion will be shared in two main directions, brought to the
fore by the structure of the equation describing the primary flux.
 Production and propagation may be
disentangled in the limit where the energy does not change as
antiprotons travel. That is not strictly correct as diffusive
reacceleration as well as adiabatic and Coulomb losses generate
a diffusion in energy space that is dicussed in Sec.~\ref{sec:energy_redist}.
The elementary process of supersymmetric antiproton production
through neutralino annihilation is discussed in Sec.~\ref{sec:production},
both in an effective MSSM and in a supergravity--inspired model. 
The two--zone propagation--diffusion model and the dependence of the primary 
antiproton flux $\Phi_{\bar{p}}^{\rm susy}$ on the propagation parameters
is described in Sec.~\ref{sec:propagation}. 
The thickness $L$ of the magnetic halo
is naively expected to be the dominant source of uncertainty for
$\Phi_{\bar{p}}^{\rm susy}$ as the larger the confinement layers, the
larger the fiducial volume where neutralino annihilations take place
and the larger the supersymmetric antiproton flux. Actually, $L$ is
combined with the diffusion coefficient $K(E)$ and the Galactic wind
velocity $V_{c}$ in order to get a precise value for the B/C ratio --
and for the  antiproton flux. 
We present the resuls for the primary flux in Sec.~\ref{sec:res_unc}, 
where we estimate the uncertainties induced by the spread of the 
diffusion--propagation parameter space. We also briefly
discuss the modifications of $\Phi_{\bar{p}}^{\rm susy}$  due 
to different choices in the dark matter distribution function, in its 
normalization  and in the core radius values. 
In Sec.~\ref{sec:toa_susy}, the comparison of the latest antiproton
measurements with the antiproton fluxes predicted in different
supersymmetric schemes will be discussed as a function of the
propagation--diffusion parameters and of the neutralino Galactic
distribution. 
Conclusions and perspectives will be presented in
Sec.~\ref{sec:conclusion}.


\section{The neutralino--induced antiprotons: the source term}
\label{sec:production}

\begin{figure*}
{\includegraphics{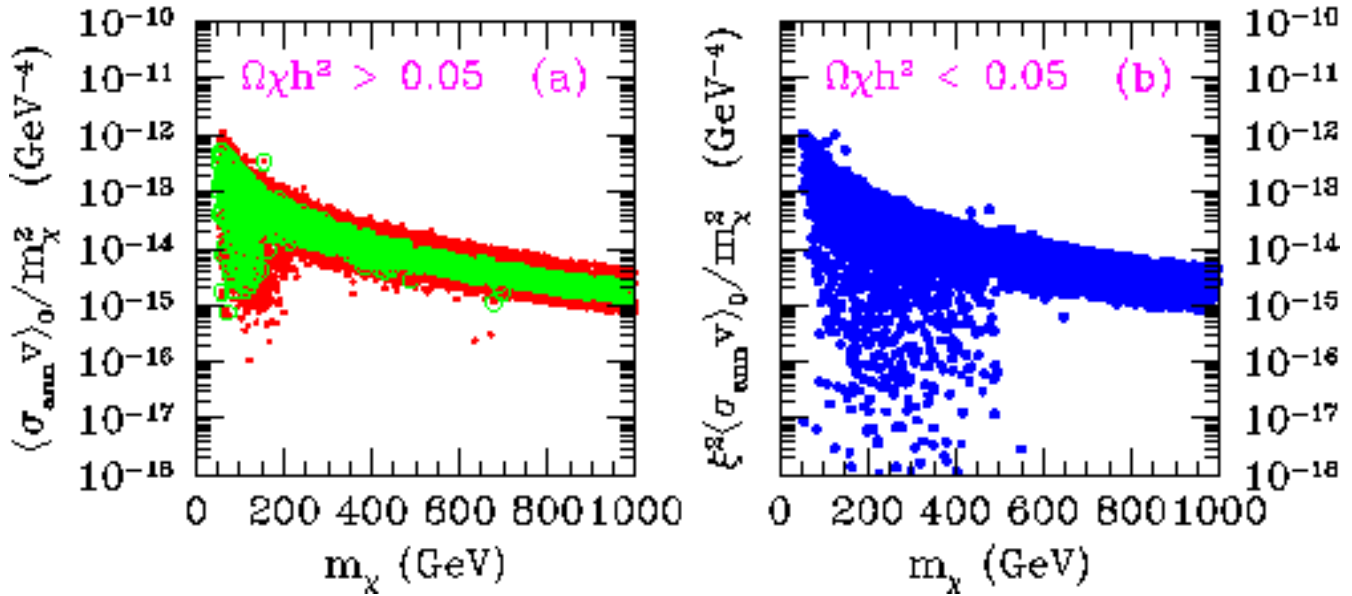}}
\caption{\label{fig:mssm_y_mchi} Scatter plot of the supersymmetric
flux factor $\Upsilon \equiv \xi^2\sigmav/m_\chi^2$ as a function of
the neutralino mass $m_\chi$, calculated in the eMSSM.  
Panel (a) refers to supersymmetric
configurations with the neutralino as a dominant dark matter component
({\em i.e.} $0.05 \leq \relic \leq 0.3$, and therefore a rescaling factor
$\xi=1$). The light (green) circles show the eMSSM configurations for 
which the neutralino relic abundance lies in the preferred range for CDM, 
as determined by the combined {\sc wmap}+2d{\sc fgrs}+Lyman--$\alpha$ analysis: $0.095 
\leq \Omega_{CDM} h^2 \leq 0.131$ \cite{cmb}. Panel (b) refers to the 
neutralino as a subdominant dark matter particle ($\relic < 0.05$).
}
\end{figure*}
\begin{figure*}
{\includegraphics{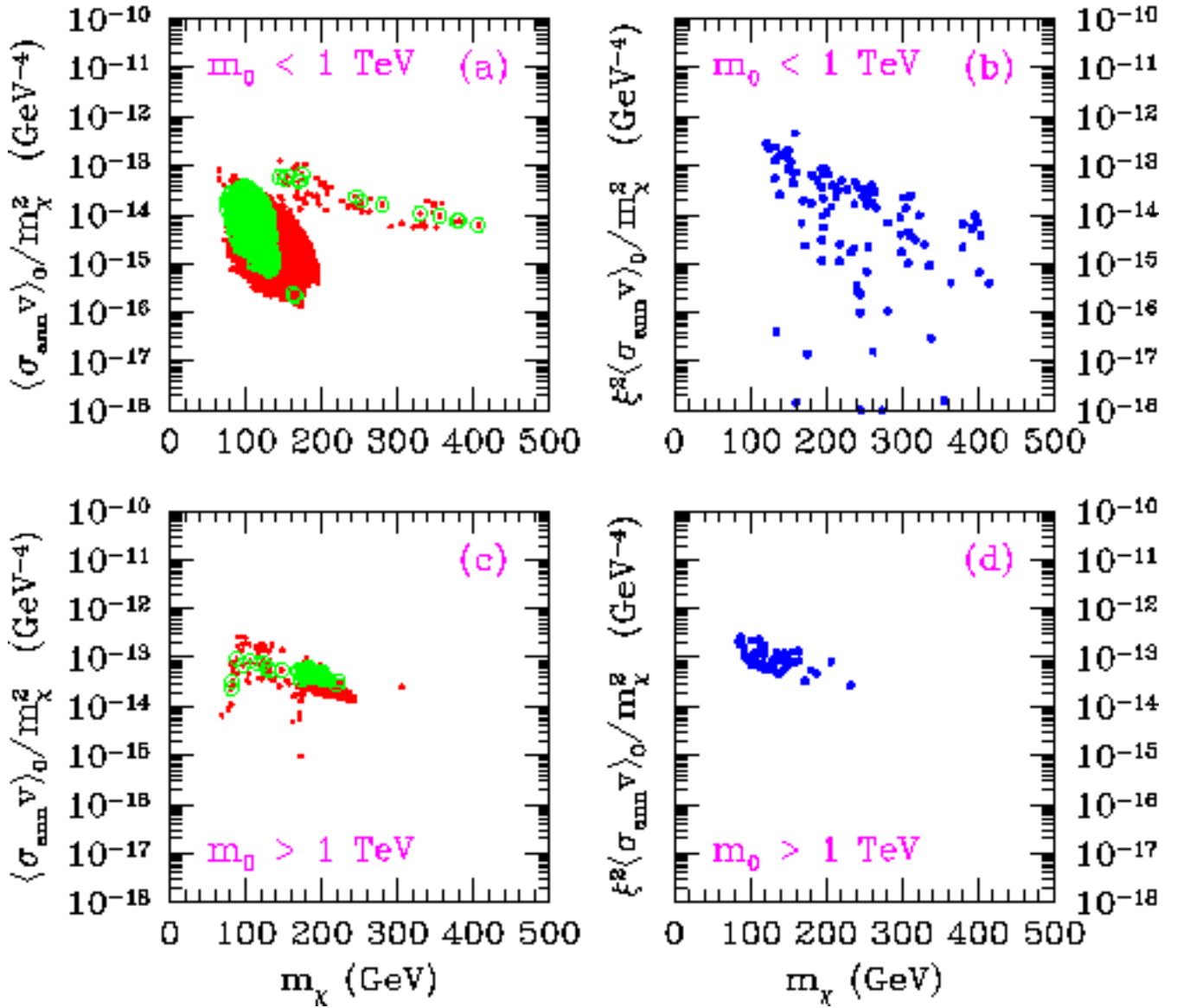}}
\caption{\label{fig:sugra_y_mchi} The same as in
Fig. \ref{fig:mssm_y_mchi}, calculated in the mSUGRA scheme. Panels (a)
and (c) refer to cosmologically dominant neutralinos ($0.05 \leq \relic \leq
0.3$); panels (b) and (d) to subdominant neutralinos ($\relic <
0.05$). The upper row (panels (a) and (b)) is obtained for the
universal soft--scalar mass $m_0$ smaller than 1 TeV (for these
models, the neutralino is mostly a bino state); the lower row (panels
(c) and (d)) refers to values of $m_0$ in excess of 1 TeV (in this
case the neutralino may have a substantial higgsino component). The
light (green) circles in panel (a) and (c) show the mSUGRA configurations for
which the neutralino relic abundance lies in the preferred range for
CDM, as determined by the combined {\sc wmap}+2d{\sc fgrs}+Lyman--$\alpha$
analysis: $0.095 \leq \Omega_{CDM} h^2 \leq 0.131$ \cite{cmb}.}
\end{figure*}
\begin{figure*}
{\includegraphics[width=1.0\textwidth]{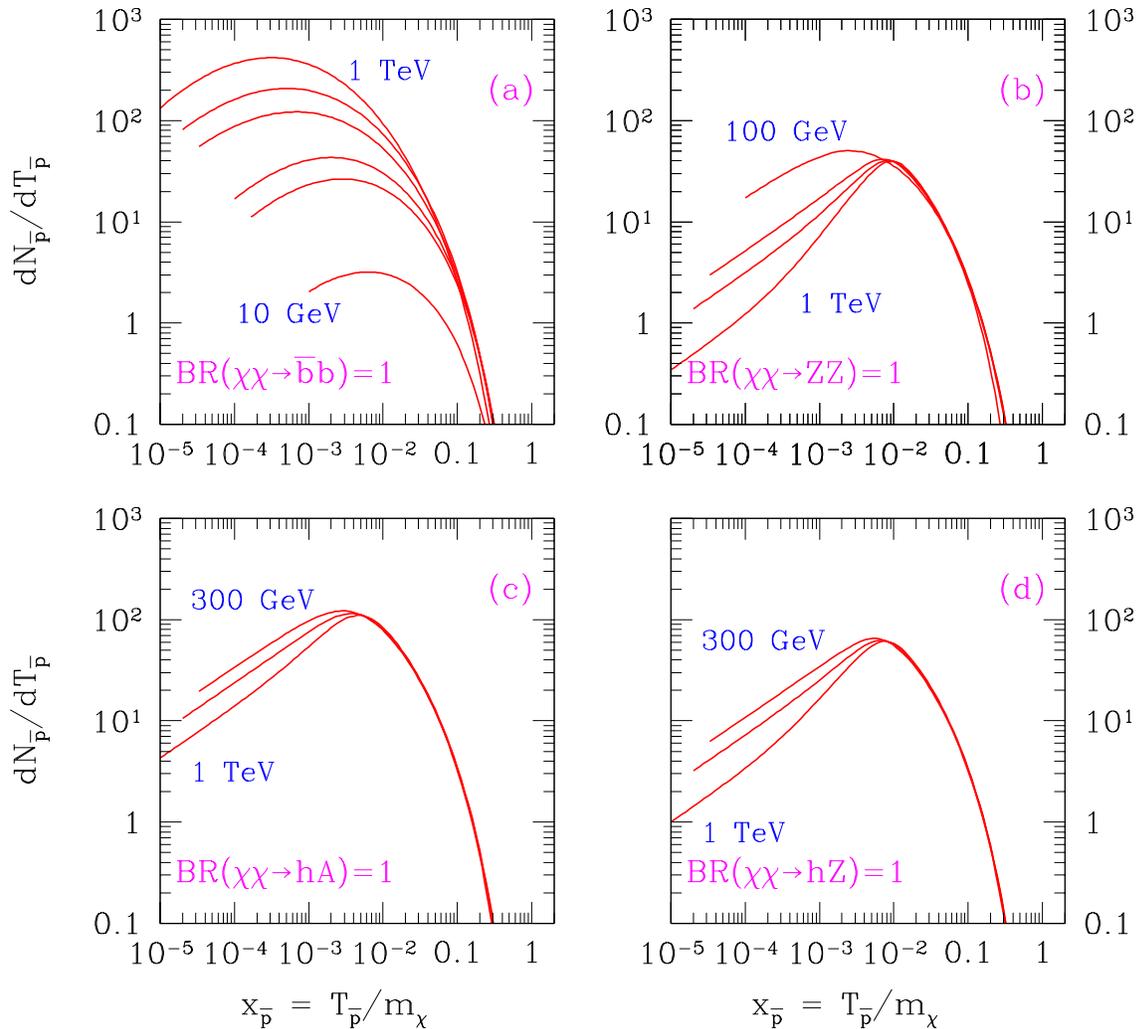}}
\vspace{-45pt}
\caption{\label{fig:g_pure}
Antiproton differential energy distribution for pure annihilation channels
as a function of the reduced kinetic energy $x_{\bar p}\equiv T_{\bar
p}/m_\chi$. Panel (a) refers to annihilation into a $b \bar b$ pair,
for neutralino masses of: $m_\chi=10,60,100,300,500,1000$ GeV (from
bottom to top); panel (b) refers to annihilation into a $ZZ$ pair, for
$m_\chi=100,300,500,1000$ GeV (from top to bottom); 
panel (c) refers to annihilation into a
scalar+pseudoscalar higgs pair $hA$, for $m_\chi=300,500,1000$ GeV
(from top to bottom), and for: $m_h=120$ GeV, $m_A=200$ GeV,
tan$\beta=10$ (ratio of higgs vev's) and $\alpha=0$ (higgs mixing
parameter); panel (d) refers to annihilation into a $hZ$ pair, for
$m_\chi=300,500,1000$ GeV (from top to bottom), and for: $m_h=120$
GeV, tan$\beta=10$ and $\alpha=0$.}
\end{figure*}
\begin{figure*}
{\includegraphics{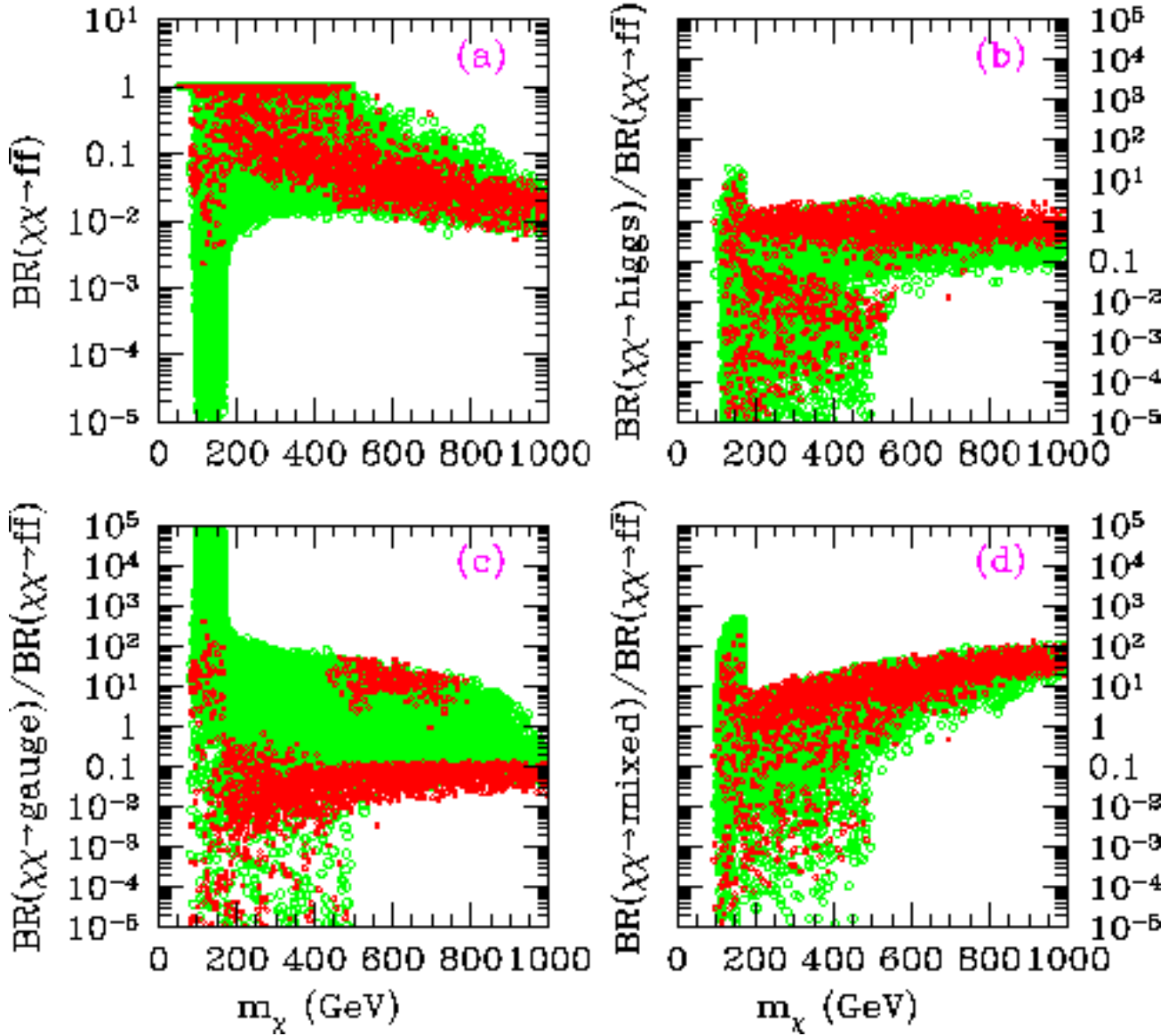}}
\caption{\label{fig:mssm_branching}Branching ratios for the neutralino
self--annihilation cross section in the eMSSM. Panel (a) shows the
amount of the branching ratio for the annihilation into a
fermion--antifermion final state ($\chi\chi\rightarrow f\bar
f$). Panels (b), (c) and (d) show the amount, relative to the $f\bar
f$ final state, of the annihilation into higgs bosons, gauge bosons
and the mixed higgs-gauge bosons final state. Dark (red) points denote
configuration with $0.05\leq \relic \leq 0.3$ (dominant relic
neutralinos). Light (green) circles indicate configuration with
$\relic < 0.05$ (subdominant relic neutralinos).}
\end{figure*}
\begin{figure*}
{\includegraphics{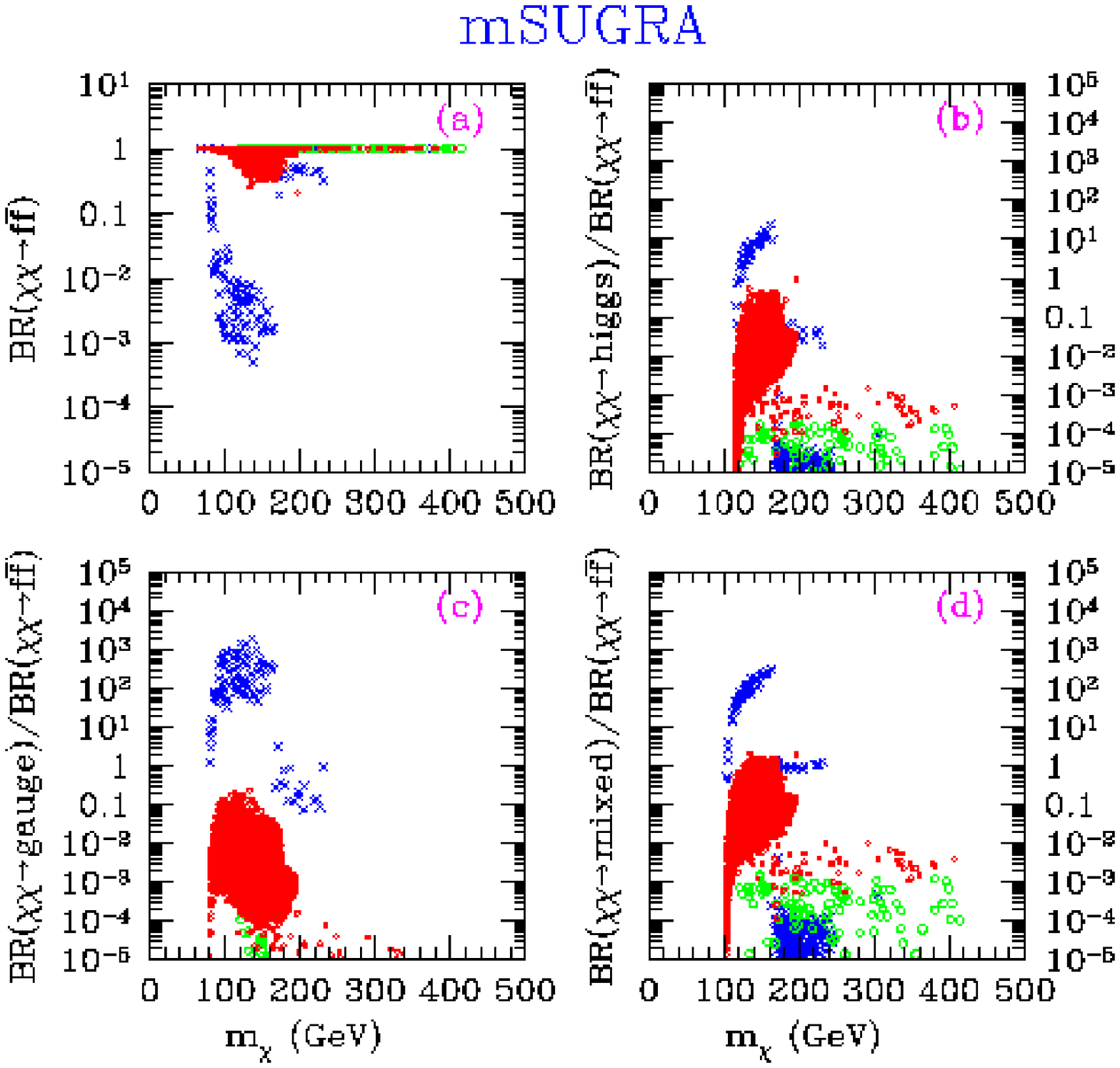}}
\caption{\label{fig:sugra_branching}The same as in
Fig. \ref{fig:mssm_branching}, calculated in the mSUGRA scheme. Dark
(red) points denote configuration with $0.05\leq \relic \leq 0.3$
(dominant relic neutralinos). Light (green) circles indicate
configuration with $\relic < 0.05$ (subdominant relic neutralinos).
Crosses (in blue) indicate the mSUGRA configurations with $m_0>1$
TeV.}
\end{figure*}

Antiprotons can be produced by self--annihilation of neutralinos in
the Galactic halo. Dark matter neutralinos may be considered almost at
rest in the Galactic frame since their average velocity is of the
order of 300~km~s$^{-1}$. They are therefore highly non--relativistic. The
production differential rate per unit volume and time is a function of
space coordinates ($r$,$z$ defined in the Galactic rest frame) and
antiproton kinetic energy $T_{\bar p}$. It is defined as:
\begin{equation}
q_{\bar p}^{\rm susy}(r,z,T_{\bar p}) =
\, \sigmav \, g(T_{\bar p})
\left( \frac{ \rho_\chi (r,z)}{m_\chi} \right)^2 ,
\label{eq:source}
\end{equation}
where $\sigmav$ denotes the average over the Galactic velocity
distribution function of the neutralino pair annihilation cross
section $\sigma_{\rm ann}$ multiplied by the relative velocity $v$,
$m_\chi$ is the neutralino mass and $\rho_\chi (r,z)$ is the mass
distribution function of neutralinos inside the Galactic halo.  Since
relic neutralinos behave as cold dark matter, their distribution has
to follow the matter density profile $\rho_{\rm DM} (r,z)$ of the
Galactic halo:
\begin{equation}
\rho_\chi (r,z) = \xi \rho_{\rm DM} (r,z)
\end{equation}
where $\xi$ parameterizes the fact that the dark halo may not be
totally made of relic neutralinos ($\xi\leq 1$).  This would be the
case when neutralinos are not responsible for the total amount of dark
matter in the Universe, {\em i.e.} when their relic abundance $\relic$
is much smaller than the measured value for $\Omega_{\rm DM}h^2$. This
is a situation which occurs in many supersymmetric models.  It is
reasonable to assume that $\xi$ has no space dependence and that is
related to the relative amount of $\relic$ with respect to
$\Omega_{\rm DM}h^2$. We will assume the standard definition:
\begin{equation}
\xi=\min(1,\relic/0.05)
\end{equation}
where we have considered that neutralinos with $\relic<0.05$ cannot be
the dominant dark matter component. 

Finally, the second term in Eq.~(\ref{eq:source}), $g(T_{\bar p})$, denotes the
antiproton differential spectrum per annihilation event, defined as:
\begin{eqnarray}
g(T_{\bar p}) &\equiv& {1 \over \sigma_{\rm ann}}
{{d \sigma_{\rm ann} (\chi \chi \rightarrow
 \bar p + X)} \over {dT_{\bar p}}} \nonumber \\ 
&=& \sum_{F}
{\rm BR(\chi\chi\rightarrow F)} \left( {dN^{F}_{\bar p}\over dT_{\bar p}} \right),
\label{eq:gE}
\end{eqnarray}
where $F$ lists the $\chi\chi$ annihilation final--state particles
which can subsequently produce antiprotons either directly
(hadronization when $F=$ quarks or gluons, or through subsequent decay
of $F$ into quarks or gluons), $BR(\chi\chi\rightarrow F)$ is the branching
ratio for the production of $F$ and $dN^F_{\bar p}/dT_{\bar p}$ denotes
the differential energy distribution of the antiprotons generated by
$F$. For details on the calculation of $g(T_{\bar p})$, see
App. \ref{sec:appendix} and Ref. \cite{antip1}.

The source term $q_{\bar p}^{\rm susy}(r,z,T_{\bar p})$ is therefore a
combination of astrophysical factors (the dark matter density profile
of the Galactic halo) and of particle physics properties (the
neutralino self--annihilation cross section and the hadronization into
antiprotons of the neutralino annihilation products). The astrophysical
and particle physics quantities are factored out and can be studied
separately. With the definitions given above, we can rewrite the antiproton 
source term as:
\begin{equation}
q_{\bar p}^{\rm susy}(r,z,T_{\bar p}) = \Upsilon
g(T_{\bar p}) \rho^2_{\rm DM} (r,z)
\label{eq:source2}
\end{equation}
where we have defined the supersymmetric flux factor $\Upsilon$ as:
\begin{equation}
\Upsilon = \xi^2 \, \frac{\sigmav}{m_\chi^2}
\label{eq:upsilon}
\end{equation}
which entirely depends on properties of supersymmetric models.

We move now to discuss each term separately.


\subsection{The Galactic distribution of dark matter}
\label{sec:profile}

For most of our discussion, we will assume that the dark matter
density distribution is described by a cored isothermal sphere. In
terms of the radial distance $r$ in the Galactic plane and of the
vertical coordinate $z$, the density profile is:
\begin{equation}
\rho_{\rm DM} (r,z) = \rho_l\,\, \frac{a^2 + R^2_\odot}{a^2 + r^2 +
z^2},
\label{eq:mass_DF}
\end{equation}
where $a$ denotes the core radius of the dark halo and $R_\odot$ is
the distance of the Sun from the Galactic center. We have set $a$= 3.5
kpc and the IAU-recommended value $R_{\odot}$ = 8.5 kpc. The value
$\rho_l$ for the total local dark matter density is determined by taking
into account the contribution given by the matter density of
Eq.~(\ref{eq:mass_DF}) to the local rotational velocity $v_{\rm rot}$
\cite{bcfs}. The value of $\rho_l$ compatible with observations ranges
from 0.18 GeV cm$^{-3}$ (for a low value of the rotational velocity,
$v_{\rm rot} = 170$ km s$^{-1}$ and a non--maximal dark halo) to 0.71
GeV cm$^{-3}$ (for $v_{\rm rot} = 270$ km s$^{-1}$ and a maximal dark
halo) \cite{bcfs}. The interval relative to the preferred value for
the rotational velocity ($v_{\rm rot} = 220$ km s$^{-1}$) is $0.30
{\rm ~GeV~cm}^{-3} \lsim \rho_l \lsim 0.47 {\rm ~GeV~cm}^{-3}$
\cite{bcfs}. Our results will be presented for $\rho_l = 0.3$ GeV
cm$^{-3}$. Since in the primary antiproton flux $\rho_l^2$ enters as a
pure normalization factor, the fluxes obtained for different values of
$\rho_l$ are easily rescaled. For instance, for $\rho_l = 0.47$ GeV
cm$^{-3}$, the antiproton fluxes would be a factor 2.45 higher than
the corresponding ones for $\rho_l = 0.3$ GeV cm$^{-3}$.

We will come back to the topic of the dark matter density profile at the
end of the paper, in Sec. \ref{sec:dm_profile}.


\subsection{Supersymmetric models}
\label{sec:susy_models}

The existence of a relic particle in supersymmetric theories arises
from the conservation of a symmetry, $R$--parity, which prevents the
lightest of all the superpartners from decaying. The nature and the
properties of this particle depend on the way supersymmetry is
broken. The neutralino can be the dark matter candidate in models
where supersymmetry is broken through gravity-- (or anomaly--)
mediated mechanisms. The actual implementation of a specific supersymmetric
scheme depends on a number of assumptions on the structure of the
model and on the relations among its parameters. This induces a large
variability of the phenomenology of neutralino dark matter. In this
paper we will consider neutralino dark matter in two different
supersymmetric schemes: a low--energy effective--theory implementation of
the Minimal Supersymmetric Standard Model (eMSSM) and a minimal
Supergravity model (mSUGRA).

The eMSSM is defined as an implementation of supersymmetry directly
at the electroweak scale, which is where the phenomenology of
neutralino dark matter is actually studied. The large number of free
parameters is reduced by a set of assumptions which are sufficient to
shape the properties of the model at the electroweak scale.  All the
relevant parameters, which set the mass scales and couplings of all the
supersymmetric particles (and of the higgs sector) are taken into
account. The free parameters are: the gaugino mass parameter $M_2$,
the higgs mixing parameters $\mu$, the ratio of the two higgs vacuum
expectation values $\tan\beta$, the mass of the pseudoscalar higgs
$m_A$, a common soft--scalar mass for the squarks $m_{\tilde q}$, a
common soft scalar mass for the sleptons $m_{\tilde l}$, a common
dimensionless trilinear--parameter $A$ for the third family ($A_{\tilde
b} = A_{\tilde t} \equiv A m_{\tilde q}$ and $A_{\tilde \tau} \equiv A
m_{\tilde l}$; the trilinear parameters for the other families are
set equal to zero). We assume the standard grand unification relation
between the $U(1)$ and the $SU(2)$ gaugino mass parameters: $M_1 = 5/3
\tan^2\theta_W M_2$. The parameters will be varied in the following
intervals: $100 {\rm ~GeV} \leq M_2 \leq 1000 {\rm ~GeV}$, $100 {\rm
~GeV} \leq |\mu| \leq 1000 {\rm ~GeV}$, $100 {\rm ~GeV} \leq m_A \leq
1000 {\rm ~GeV}$, $100 {\rm ~GeV} \leq m_{\tilde q}, m_{\tilde l} \leq
3000 {\rm ~GeV}$, $1 \leq \tan\beta \leq 50$ and $-3 \leq A \leq 3$.

A different approach is to embed supersymmetry in a supergravity
scheme with boundary conditions at some critical high energy scale,
such as the grand unification (GUT) scale, and keeping the number of
free parameters and assumptions minimal. This is our
mSUGRA.  In this class of models we consider gauge coupling constant
unification at the GUT scale. In addition, all the mass parameters in
the supersymmetric breaking sector are universal at the same GUT
scale. The low--energy sector of the model is obtained by evolving all
the parameters through renormalization group equations from the GUT
scale down to the electroweak scale: this process also induces the
breaking of the electroweak symmetry in a radiative way. This model is
very predictive, since it relies only on very few free parameters, but
at the same time it has a very constrained phenomenology at
low--energy. It also appears to be quite sensitive to some standard
model parameters, like the mass of the top and bottom quarks ($m_t$
and $m_b$) and the strong coupling constant $\alpha_s$. In this class
of models there are four  free parameters: the universal gaugino mass
parameter $M_{1/2}$ at the GUT scale, the universal soft--scalar mass
parameter for both the sfermions and the higgses $m_0$ at the GUT
scale, a common trilinear coupling for the third family at the GUT
scale $A_0$ and $\tan\beta$.  The parameters will be varied in the
following intervals: $50 {\rm ~GeV} \leq M_{1/2} \leq 1000 {\rm
~GeV}$, $0 \leq m_0 \leq 3000 {\rm ~GeV}$, $1 \leq \tan\beta \leq 50$
and $-3 \leq A_0 \leq 3$. The standard model parameters $m_t$, $m_b$
and $\alpha_s$ are varied inside their $2\sigma$ allowed ranges.


\subsection{The supersymmetric flux factor $\Upsilon$}
\label{sec:flux_factor}

The flux factor $\Upsilon$ defined in Eq.~(\ref{eq:upsilon}) acts as a
normalization factor for the antiproton flux and is a purely
supersymmetric term. It depends on the mass and couplings of
neutralinos in the supersymmetric framework under study. In
Fig. \ref{fig:mssm_y_mchi} we show the flux factor as a function of the
neutralino mass for a scan of the eMSSM. Fig. \ref{fig:sugra_y_mchi}
reports the case for the mSUGRA scheme. We show the values of
$\Upsilon$ separately for the case of comsologically dominant ($0.05
\leq \relic \leq 0.3$, Fig. \ref{fig:mssm_y_mchi}a and
Figs. \ref{fig:sugra_y_mchi}a,c) and subdominant ($\relic < 0.05$,
Fig. \ref{fig:mssm_y_mchi}b and Figs. \ref{fig:sugra_y_mchi}b,d) relic
neutralinos. Among the cosmologically relevant ones, we also show the
configurations which yield $\relic$ inside the preferred range for
CDM, as determined by the combined {\sc wmap}+2d{\sc fgrs}+Lyman--$\alpha$
analysis: $0.095 \leq \Omega_{CDM} h^2 \leq 0.131$ \cite{cmb}. The
results in the eMSSM show that the upper values of $\Upsilon$ are
around $10^{-12}$ GeV$^{-4}$ for neutralino masses close to the
experimental lower bound (around 50 GeV) and then decrease below
$10^{-14}$ GeV$^{-4}$ for $m_\chi \sim 1$ TeV.

In the case of dominant relic neutralinos, the interval of values for
$\Upsilon$ is restricted, at all masses: in order to have values of
$\relic$ which fall in the cosmologically relevant range, the
annihilation cross section integrated from freeze--out down to the
present time must be inside the interval $3\cdot 10^{-11} {\rm
~GeV}^{-2} \lsim \sigmavint \lsim 2\cdot 10^{-10} {\rm ~GeV}^{-2}$. We
remind that the relic abundance depends on $\sigmavint$ ($\relic
\propto \sigmavint^{-1}$). This cross section, due to the
non--vanishing temperature in the early Universe, may differ quite
substantially from the zero--temperature cross section $\sigmav$, which is
instead relevant for the antiproton signal. Usually a correlation
between $\sigmavint$ and $\sigmav$ is present when the
zero--temperature $\sigmav$ is large; on the contrary, when $\sigmav$
is small, temperature corrections in the early Universe induce
$\sigmavint$ to deviate, also sizeably, from $\sigmav$. This
difference in the two cross sections is responsible for the band of
values of $\Upsilon$ shown in Fig. \ref{fig:mssm_y_mchi}a.

When the neutralino relic abundance is low, such that neutralinos are
not the dominant component of dark matter, $\Upsilon$ acquires an
additional dependence on $\relic$ through the rescaling factor
$\xi^2$. This is shown in Fig. \ref{fig:mssm_y_mchi}b. The effect of
$\xi^2$ is obviously to reduce $\Upsilon$: the lower the relic
abundance, the smaller $\xi$ and then the flux factor. The lowest
points in Fig. \ref{fig:mssm_y_mchi}b are the ones with lower values of
$\relic$. These configurations, even though they give a large
$\sigmav$ (low $\relic$ has large values of $\sigmavint$, and in this
case $\sigmavint \propto \sigmav$), nevertheless they have a low flux
factor because they are under--abundant. This implies that largely
subdominant relic neutralinos are likely to provide (almost)
undetectable antiproton fluxes. This is somewhat at variance with the
case of direct detection: the difference arises from
the fact that the antiproton signal (as well as the other Galactic
signals) depends quadratically on the dark matter density (and
henceforth on the rescaling factor) while for direct detection the
dependence is linear and the suppression is much milder
\cite{direct0}.

The situation of mSUGRA is shown in Fig. \ref{fig:sugra_y_mchi}. In
this case, the largest values of the flux factor $\Upsilon$ are about
an order of magnitude lower that in the eMSSM: $\Upsilon \lsim
10^{-13}$ GeV$^{-4}$ for light neutralinos. This is a consequence of
the properties of neutralinos in this constrained type of models:
neutralinos turn out to be mainly gauginos and their couplings,
especially to higgs bosons which require a mixed
higgsino--gaugino neutralino content, are in general smaller than 
in some sectors of the eMSSM. The lower panels in
Fig. \ref{fig:sugra_y_mchi} show the situation in a sector of the
mSUGRA scheme where the soft scalar masses are large: $m_0>1$ TeV
\cite{bere,focus}. In this sector, the neutralino may acquire a
non--vanishing higgsino component \cite{bere,focus}, as a consequence
of the radiative electroweak symmetry breaking, and their couplings to
higgs bosons are enhanced \cite{bere}: the consequence on the flux
factor is in fact a mild enhancement, up to values of $\Upsilon$
around $3-4 \cdot 10^{-13}$ GeV$^{-4}$, closer to the eMSSM upper
values.

\subsection{The differential antiproton spectrum $g(T_{\bar p})$}
\label{sec:spectrum}

\begin{figure}[t]
\vspace{-30pt}
{\includegraphics[width=\columnwidth]{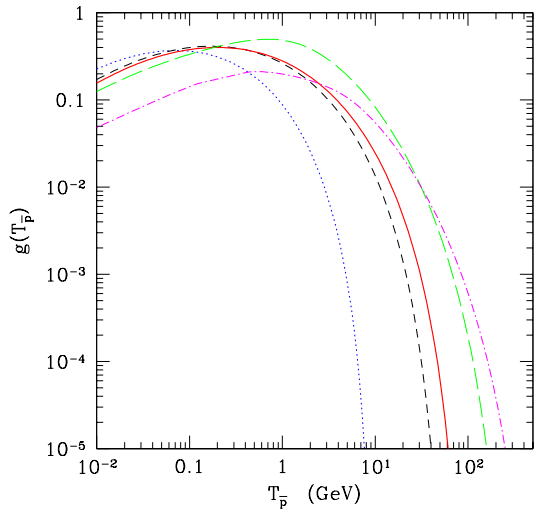}}
\vspace{-20pt}
\caption{\label{fig:g_E} Representative differential antiproton
spectra per annihilation event $g(T_{\bar p})$ from neutralino
self--annihilation, as a function of the antiproton kinetic energy
$T_{\bar p}$. The different curves refer to different neutralino
masses: $m_\chi$=10 (dotted [blue]), 60 (short dashed [black]), 100
(solid [red]), 300 (long dashed [green]) and 500 GeV (dot--dashed
[magenta]). The spectra are selected from our sample of the eMSSM,
except the one for $m_\chi = 10$ GeV which refers to an eMSSM without
grand--unification gaugino universality \cite{light,wmap}. All the
spectra refer to neutralinos with $\relic=0.1$ and large values of the
flux factor $\Upsilon$. The spectrum for $m_\chi =100$ GeV is the
reference spectrum for the analysis of the astrophysical properties in
the next Sections.}
\end{figure}

Let us move now to the discussion of differential spectra of
antiprotons which are produced by neutralino annihilation. The
capability of producing antiprotons depends on the possibility for
neutralinos to produce quarks or gluons, either directly or through
decay of their annihilation products: quarks and gluons will then
hadronize and eventually produce antiprotons. We have modelled the
hadronization process by using the PYTHIA Monte Carlo \cite{pythia}. 
The neutralino annihilation is calculated analytically as described in
Ref. \cite{anni}. Neutralino annihilation occurs at rest in the
Galactic frame, and the different final states which are open depend
therefore on the neutralino mass. The annihilation may proceed through
the following channels: production of a fermion pair; production of
$WW$ and $ZZ$; production of a higgs pair; production of a higgs
together with a gauge boson (which can be the $Z$ boson or the $W$
depending whether the higgs is neutral or charged). Apart from the
direct production of quarks or gluons, the decay chain of the
annihilation products until a quark is produced is calculated
analytically. At this stage, the antiproton differential flux is
obtained from the MC modelling. More details are given in App.
\ref{sec:appendix}.

A sample of $\bar p$ spectra for the four types of neutralino
annihilation final states is shown in Fig. \ref{fig:g_pure}, for
different values of the neutralino mass: panel (a) shows the spectra
calculated for annihilation into a pure $b \bar b$ state; panel (b)
refers to annihilation into a pure $ZZ$ state; panel (c) refers to an
annihilation into a higgs pair, where the scalar higgs has mass of
$m_h=120$ GeV, the pseudoscalar higgs mass is $m_A=200$ GeV, for
$\tan\beta=10$ and for a vanishing value of the higgs mixing parameter
$\alpha$; panel (d) refers to annihilation into a $hZ$ pair, for:
$m_h=120$ GeV, $\tan\beta=10$ and $\alpha=0$. Fig. \ref{fig:g_pure}
shows the dependence of the antiproton spectra on  the production
energy, fixed by the neutralino mass. For instance, in panel (a) the
antiprotons are produced by the hadronization of $b$ quarks injected
at the energy given by the neutralino mass; in panel (b), antiprotons
are produced by quarks produced by the decay of $Z$ bosons in motion
with respect to the neutralino rest frame: a Lorentz boost on the
hadronization spectra is therefore operative in shifting the fluxes
to larger kinetic energies.

Spectra like the ones shown in Fig. \ref{fig:g_pure} are used to
calculate the differential spectra $g(T_{\bar p})$.  However, as it is
clear from Eq.~(\ref{eq:gE}), we also need to know the values of the
branching ratios of each neutralino annihilation final state. The
branching ratios will weight the different differential spectra, like
the ones shown in Fig.~\ref{fig:g_pure}. An example of branching
ratios for neutralino annihilation is given in
Fig.~\ref{fig:mssm_branching} for the eMSSM scheme, and in
Fig.~\ref{fig:sugra_branching} for the mSUGRA models.

In the eMSSM, we notice that the annihilation in fermions may
be sizeable and dominant for masses lower than 500 GeV. The
two higgses final state is usually of the order or lower than
the $f \bar f$ final state, while the gauge bosons final state may
dominate, except for very large neutralino masses. The mixed gauge+higgs
boson final state tends to be dominant at very large neutralino masses.

In the case of mSUGRA models, since the neutralino tends to be a
gaugino which couples effectively to fermions through sfermion
exchange, the $f \bar f$ final state usually dominates. A relevant
production of final states other than fermions, especially gauge
bosons, occurs in the large sfermion masses regime ($m_0>1$ TeV),
where sfermion-exchange is suppressed by the large sfermion mass and
at the same time a higgsino component for the neutralino arises: this
facilitates both the coupling to higgses and to gauge bosons.

The final result of this analysis is the calculation of realistic
antiproton differential spectra for neutralino annihilation. Some
representative examples are shown in Fig.~\ref{fig:g_E}, for different
values of the neutralino mass. All the spectra refer to neutralinos
selected to have $\relic = 0.1$ and large values of the flux factor
$\Upsilon$ (close to the upper values of Fig.~\ref{fig:mssm_y_mchi},
for each mass). All these spectra properly take into account all the
ingredients discussed in this Section: the hadronization spectra and
the annihilation branching ratios. The spectrum for $m_\chi =100$ GeV
is the reference spectrum for the analysis of the astrophysical
properties in the next Sections.  The spectra shown in
Fig.~\ref{fig:g_E} are selected from our sample of the eMSSM, except
the one for $m_\chi = 10$ GeV which refers to an eMSSM without the
grand--unification condition between the gaugino mass parameters $M_1$
and $M_2$ \cite{light,wmap}. In this class of models the neutralino
can be as light as a few GeV \cite{light,wmap}, in contrast to the
standard eMSSM, where LEP constraints imply a lower bound on the
neutralino mass of about 50 GeV. For completeness, we have therefore
included also the representative spectrum for $m_\chi=10$ GeV, in
order to illustrate the effect of propagation on the primary flux from
light neutralinos. However, a complete study of the eMSSM without
grand--unification gaugino universality is beyond the scope of this
paper.

Now that we have discussed the source term, we proceed to the second
step of the calculation: the study of how these antiprotons diffuse
and propagate in the Galaxy and in the Solar System. The result of
this analysis will be the interstellar and top--of--atmosphere (TOA)
fluxes of primary antiprotons.


\section{Diffusion and propagation in the Galaxy}
\label{sec:propagation}

The propagation of cosmic rays in the Galaxy has been considered in
the framework of a two-zone diffusion model, which has been described
at length in Refs. \cite{PaperI,PaperII,revue}. Here we only remind
the main features of this model, and refer to the above-mentioned
papers for all the details and motivations. We also present in detail
the quantitative dependence of the secondary and primary signal on the
propagation parameters.


	\subsection{The framework}

The disk of the Galaxy is described as a thin disk of
radius $R=20$ kpc, which contains the interstellar gas with a surface density
$\Sigma = 2h n_{\rm ISM}$ with $h=100$ pc and $n_{\rm ISM} = 1$ cm$^{-3}$.
It is embedded in a thicker diffusion halo, supposed to have a
cylindrical shape with the same radius $R$ as
the disk and height $L$ which is not well known. The matter density
is much lower in the diffusion halo
so that spallations (rate $\Gamma\equiv 2hn_{\rm ISM} \sigma$
yielding the secondary species) of the charged
nuclei occur only in the disk.
Moreover, the standard sources also happen to be located in the disk.

The spatial diffusion of cosmic rays is assumed to occur uniformly in
the whole (disk and halo) diffusion volume, with the same strength.
The corresponding diffusion coefficient has been defined as $K(E) =
K_0 \beta ({\cal R}/1\, {\rm GV})^\delta$, where ${\cal R}$ stands for
the particle rigidity and $K_0$ and $\delta$ are free parameters of
the model.  We also consider the possibility that a Galactic wind
blows the particles away from the disk in the $z$ vertical direction,
with a constant speed $V_c$. It induces an adiabatic dilution of the
energy of the particles in the disk due to the sudden change in
$V_c$. Several other processes modify the antiproton energy distribution:
ionization losses when interacting with the neutral interstellar
matter, or from Coulomb losses in a completely ionized plasma,
dominated by scattering off the thermal electrons. To end with,
minimal reacceleration on random hydrodynamic waves, {\em i.e.} diffusion in
momentum space, described by a coefficient $K_{pp}$ related to the
spatial diffusion $K(E)$, is inevitable \cite{Ptuskin}. This process
is assumed to occur only in the disk and is related to the velocity of
disturbances in the hydrodynamical plasma $V_A$, called Alfv\'en
velocity. In summary, our diffusion model has five free parameters
$K_0, \delta, L, V_c, V_A$ which describe the minimal number of
physical effects thought to have some role in antiproton propagation.

The sets of diffusion parameters were constrained in a previous work
\cite{PaperI} (see also Ref. \cite{PaperIb}) by analysing stable nuclei
(mainly by fitting the boron to carbon ratio B/C). The values we
obtained were also shown to be compatible with the observed secondary
antiprotons \cite{PaperII} and the flux of radioactive isotopes
\cite{beta_rad}.  However, in a first step we will disregard these
constraints: in order to clarify which of the propagation parameters
are important if one wants to compare any possible primary component
to the background (secondaries), we study the effect of each parameter
on the signal and the background.  Only in a second step this
additional information on the constraints is used to conclude about
the variation that will result in the primary supersymmetric signal
(see Sec.~\ref{Sec:propagated_flux}).


\subsection{Solutions for primary and secondary antiprotons}

We are interested in the cosmic ray antiproton flux:
\begin{equation}
      \Phi^{\bar{p}}(r,z,E) = \frac{v_{\bar{p}}}{4\pi} N^{\bar{p}}(r,z,E)\;\;.
\end{equation}
It is related to the differential density $N^{\bar{p}}(r,z,E)\equiv
dN^{\bar{p}}(r,z,E)/dE$ which satisfies the steady-state diffusion
equation. The general procedure to solve this equation as well as
references for detailed derivation is given in
App.~\ref{sec:appendix2}. At variance with the solutions already
presented elsewhere, it proves to be useful, as suggested by the study
of the spatial origin of secondary and primary CRs (see
Ref. \cite{origin} for what is ment by ``spatial origin"), to
introduce the quantities:
\begin{eqnarray}
	r_{\rm w}  &\equiv& \frac{2K(E)}{V_c} \;, \\
        r_{\rm sp} &\equiv& \frac{K(E)}{h \Gamma_{\rm inel}(E)} \;.
\end{eqnarray}
Since many configurations of  $K(E)=K_0\beta {\cal R}^\delta$ and $V_c$
lead to the same $r_{\rm w}$ and $r_{\rm sp}$, these new parameters
avoid automatically a useless discussion about many degenerate values
of the diffusion  coefficient  and make the dependence over the
important parameters more evident in formulae (the 
physical meaning
of these new parameters is explicited below, see also Ref. \cite{origin}).

The solutions are given below discarding energy redistributions
(see App.~\ref{sec:energy_redist} for the procedure to include them).
Energetics are not the dominant effects so it is interesting to
focus on the analytical formulae obtained in that case.

\paragraph{The primaries}
Let us first inspect the primaries: the source term is given by
$q_{\bar p}^{\rm susy}(r,z,T_{\bar{p}})$ described by
Eq.~(\ref{eq:source}) and discussed in detail in
Sec.~\ref{sec:production}. Primary antiprotons are produced throughout
the whole diffusive halo, which is embedded in the dark matter halo.
An advantage when energy losses and gains are discarded is that the
solution can be recast as (see App. \ref{sec:appendix2})
\[
N^{\bar{p},{\rm prim}}(r=R_{\odot},z=0,E)= 
{\cal E}^{\rm prim}_{\rm source}(E)\times
S^{\rm prim}_{\rm astro}(R_{\odot},0,E)
\]
where the {\em elementary} source term (spectrum from a point source)
given by
\begin{equation}
{\cal E}^{\rm prim}_{\rm source}(E)\equiv \Upsilon g(T_{\bar p})
\end{equation}
can be separated from the astrophysical part
\begin{eqnarray}
   S^{\rm prim}_{\rm astro}(R_{\odot},0,E)\equiv
    \sum_{i=1}^{\infty} \Pi_i(E,R_{\odot})
\left\{ \int_0^RJ_0(\zeta_ir/R)\right. \nonumber \\
\left. \int_{-L}^Le^{-z/r_{\rm w}}\frac{\sinh[S_i(L-z)/2]}
{\sinh[S_iL/2]}\times
w\left(r,z\right) \right\} dz r dr.
\label{Eq:sol_primaires}
\end{eqnarray}
In the above equation, $w(r,z)$ is the {\em effective} spatial distribution of
the primary sources [e.g. $\rho^2_{\rm DM}(r,z)$
for supersymmetric particles and $\rho_{\rm DM}(r,z)$ for evaporating
primordial black holes (PBH)].
We have defined
\begin{equation}
\Pi_i(E)\equiv\frac{2}{A_i^{\bar{p}}(E) R^2J_1^2(\zeta_i)}\times
J_0(\zeta_iR_\odot/R)\;\;.
\end{equation}
We also use
\begin{equation}
     A_i^{\bar p}(E)= K(E)\left\{ 2r_{\rm sp}^{-1}(E) +2r_{\rm w}^{-1}(E) +
       S_i \coth \left(\frac{S_iL}{2} \right)\right\}
\label{ai_reduit}
\end{equation}
and
\begin{equation}
        S_i = \sqrt{4r_{\rm w}^{-2}(E)+4\zeta_i^2/R^2}\;\;.
\label{si_reduit}
\end{equation}
The functions $J_0$ and $J_1$ are respectively the Bessel functions of 0-th
and first order, and $\zeta_i$ is the i-th zero of $J_0$. The superscript
$^{\bar{p}}$ in $ A_i^{\bar p}(E)$  indicates that the term $r_{\rm 
sp}$ should be evaluated for the antiproton destruction rate
and at the parent rigidity.

Compared to
Eq.~(\ref{eq:source}) that describes the supersymmetric source term at each
position $(r,z)$, we isolated in the new term 
${\cal E}^{\rm prim}_{\rm source}(E)$
the only required information about the production process.
For antiprotons produced by neutralino annihilations, 
the flux factor $\Upsilon$ (see Sec.~\ref{sec:flux_factor}) and the 
{\em elementary} spectrum $g(T_{\bar p})$ (see Sec.~\ref{sec:spectrum}) 
are fully described by the properties of the supersymmetric 
and hadronization models. As a result,
$S^{\rm prim}_{\rm astro}(R_{\odot},0,E)$ is solely 
dependent on the propagation properties and the effective
spatial source distribution $w(r,z)$. This function
is all we need in order to discuss the propagation uncertainties
on the primary fluxes, {\em i.e.} the signal detected at Solar location
$R_{\odot}$.

\paragraph{The secondaries.}
The secondaries are produced from proton sources distributed according to the
spatial supernova remnant distribution in the thin disk $2h\delta(z)q(r)$.
These protons are first propagated, leading to an equilibrium distribution
$N^p(r,z,E)$, that in turn produces secondary antiprotons when it interacts
with the interstellar gas. Compared to primaries, secondaries diffuse twice.
Actually, it is not possible, strictly speaking, to isolate an elementary
source term as for primaries (we skip the details, but the interested reader
can inspect the structure of equations in Ref. \cite{PaperII}). However,
it is possible to overcome this shortcoming. Antiprotons are only produced by
protons that are beyond the threshold of 7~GeV; 
in the term $A_i^p$ that originally appears in
the secondary solution (see e.g. Ref. \cite{PaperII}) -- and 
that prevents from this separation --, one can neglect
spallations and convection (high energy regime) and approximate
\[
A_i^p\approx K(E)\times 2(\zeta_i/R) \coth
(\zeta_iL/R)\;\;.
\]

It is then possible to recast the various terms entering the solution
in order to obtain a formula that (as for
primaries) isolate the dependence on the propagation terms:
\begin{equation}
N^{\bar{p},\;sec}(r=R_{\odot},0,E) \approx 
{\cal E}^{\rm sec}_{\rm source}(E)\times
B^{\rm sec}_{\rm astro}(R_{\odot},0,E)\;\;.
\end{equation}
The corresponding terms are
\begin{equation}
{\cal E}^{\rm sec}_{\rm source}(E_{\bar{p}})\equiv
\int_{E_{\rm thresh}}^{\infty}\frac{Q(E_p)}{K(E_p)}
\frac{d\sigma(E_p,E_{\bar{p}})}{dE_p} dE_p
\end{equation}
and
\begin{eqnarray}
B^{\rm sec}_{\rm astro}(R_{\odot},0,E)\equiv \sum_{i=1}^{\infty}
\frac{\Pi_i(E,R_{\odot})}{2(\zeta_i/R) \coth(\zeta_iL/R)} \nonumber \\
\cdot \left\{ \int_0^RJ_0(\zeta_ir/R) 
2hq^{\rm disk}(r) \right\} rdr\;.
\label{Eq:sol_secondaires}
\end{eqnarray}
This approximate solution is only used to estimate the 
sensitivity of the fluxes
to the diffusion parameters. We go back to the 
full solution (see Ref. \cite{PaperII} and App.~\ref{sec:appendix2} 
for more references) when the final results are presented.

\paragraph{Sensibility to the propagation parameters.}
With the quantities defined above, it is straightforward to
evaluate primaries and secondaries fluxes ``as if" the elementary
production processes were the same (to focus on the astrophysical
uncertainties). This defines the relative sensitivity to the propagation
parameters, and it is merely the ratio of  the {\em astrophysical} part of
the signal $S$ to the background $B$:
\begin{equation}
      {\cal S}ens{\rm [Par]}\equiv
      \frac{S^{\rm prim}_{\rm astro}(R_{\odot},0,
      E)}{B^{\rm sec}_{\rm astro}(R_{\odot},0,E)}\;.
      \label{eq:sens}
\end{equation}
This ratio is likely to depend on the propagation parameters,
in first place because primary sources are located in
the whole diffusive halo, whereas secondary sources are induced
spallatively in the thin disk only.

We now investigate how the primary flux
$S^{\rm prim}_{\rm astro}$, secondary flux $B^{\rm sec}_{\rm astro}$
and the relative sensitivity ${\cal S}ens$ depend on the propagation
parameters $K(E)$, $r_{\rm w}$, $r_{\rm sp}$,
$L$, $R$, $R_{\odot}$ and on the effective source distribution $w(r,z)$.
This discussion will be general and apply to any primary species.
It is discussed below for the case of supersymmetric primaries,
but we will also plot (but not comment) the results
for PBH antiprotons.


	\subsection{Evolution of fluxes with astrophysical parameters}
	\label{sec:error_primary}
We now review each one of the above parameters, starting with the diffusion
coefficient $K(E)=K_0 \beta {\cal R}^\delta$. This parameter induces
both a change in the normalization -- through $K_0$ and 
only in the high energy regime -- and in the energy dependence
(through ${\cal R}^\delta$). At sufficiently high energy (above a few 
tens of GeV), $r_{\rm w}, r_{\rm sp}\gg 1$ and $A_i$ and $S_i$ become
independent on $E$, so that the sole energy dependence
$1/K(E)$ is factored out of $\Pi_i(E,R_{\odot})$, {\em i.e.} of the 
Bessel sums. As a result, the quantity ${\cal S}ens$ is insensitive
to the choice of $K(E)$, whatever the value of the other parameters. 
There is one subtlety left: the secondary
{\em elementary} production  ${\cal E}^{\rm sec}_{\rm source}(E)$, as defined
above, contrarily to the primary's one, is not fully {\em elementary}, 
because it does depend on the value of $K(E_p)$ above 7 GeV. However,
as we will see later, all propagation parameters are designed
to have about the same $K(E)$ at 100 GeV, so that the quantity
$N^{\bar{p},\;prim}/N^{\bar{p},\;sec}(r=R_{\odot},0,E)$ is 
eventually not that sensitive to this parameter.

\subsubsection{The diffusive halo size $L$ and the radius $R$ of the Galaxy}
These parameters are related to the escape probability from the confinement
volume (the magnetic halo of the Galaxy). The larger $L$ and $R$, the greater
the probability for particles emitted in remote sources to reach us. 
Actually, the side boundary
plays almost no role for several reasons.
First, escape is driven by the closest boundary, which is
the one at $z=\pm L$ as $L$ is likely to be smaller than $R$;
second the source distribution is peaked near the Galactic
center and decreases to very small values at large radii (see Ref. \cite{origin} 
for more details). 
\begin{figure}[t]
\centerline{
\includegraphics*[width=\columnwidth]{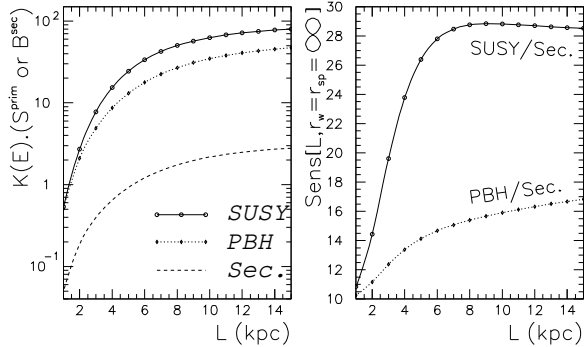}}
\caption{This plot displays the quantities 
$K(E)\times S^{\rm prim}_{\rm astro}$ and $K(E)\times
B^{\rm sec}_{\rm astro}$, see Eqs.~(\ref{Eq:sol_primaires})
and~(\ref{Eq:sol_secondaires}) 
(left panel) and ${\cal S}ens$ defined by Eq.(\ref{eq:sens}) (right
panel), as a function of the propagation parameter $L$ ($r_{\rm w}=
r_{\rm sp}=\infty$, {\em i.e.} no wind, no spallations) for an isothermal 
profile. Two cases have been considered for the primary signal. 
The curve labelled  supersymmetric corresponds to an effective
source term proportional to $w(r,z)=\rho_{DM}^2$, whereas the curve 
{\sc pbh} corresponds to
a source term proportional to $w(r,z)=\rho_{DM}$, as for primordial black holes.}
\label{effet_L}
\end{figure}
Hence, for $L\lesssim 5$~kpc, setting $R=20$~kpc or $R=\infty$ leaves
 $S$ and $B$ unchanged.  The enhancement of fluxes with $L$ can be seen
in Fig.~\ref{effet_L} (we use here and in other figures the isothermal
profile for the dark matter distribution), showing ${\cal
S}ens[L,r_{\rm w}=\infty,r_{\rm sp}=\infty]$ as a function of $L$ (we
limit the discussion to the supersymmetric case, but the reader can
straightforwardly conclude for PBH's).  For small $L$, only the
sources very close to the Solar neighboorhood contribute and, as the
dark matter source distribution is normalized to $1$, the supersymmetric and PBH
cases yield the same value.  As $L$ increases, escape is less
efficient and more sources (secondary or primary) effectively
contribute to the signal.  This enhancement is more important for
primary than for secondaries, as the effective number of sources
increases respectively as $L^3$ (volume distribution) and $L^2$
(surface distribution). In the case of primaries, part of the
enhancement is also due to the mere fact that the number of sources
within the diffusive box increases with $L$ (we remind that the
sources from the dark halo to be propagated are those enclosed inside
the diffusive box, see Ref. \cite{pbar_pbh}).  Both effects are
responsible for the evolution of ${\cal S}ens$.  For $L\gtrsim 5$~kpc,
no further significant enhancement is observed, as the bulk of the
primary sources (the core radius of the dark matter distribution) is
then almost entirely enclosed in the diffusive halo.  To be
quantitative, ${\cal S}ens$ is increased by a factor 3 for $L=15$ kpc
compared to $L=1$ kpc. Notice that the quantity $(K(E)\times S^{\rm
prim}_{\rm astro})$ plotted on the left panel of
Figs. \ref{effet_L},\ref{effet_wind},\ref{effet_spal} does not depend
on $K(E)$.  To understand this property it is sufficient to look at
the expression for $S^{\rm prim}_{\rm astro}$ in
Eq. (\ref{Eq:sol_primaires}).

\subsubsection{The Galactic wind $V_c$ through $r_{\rm w}$}
At high energy (generally a few tens of GeV), propagation is
dominated by diffusion.
At low energy convection may become the most efficient process
(parameter $r_{\rm w}\approx 1$, see Refs. \cite{Jones78,origin}) and it may
compete with $L$ for escape.
The effect of convection is to blow the particles away from the disk,
leading to an effective size of the diffusive halo $L^\star \sim r_{\rm w}$.
There is a difference with the effect of $L$ as the decrease of $r_{\rm w}$ 
does not lead to a decrease of the number of primary sources enclosed in 
the diffusive volume.
\begin{figure}[t]
\centerline{
\includegraphics*[width=\columnwidth]{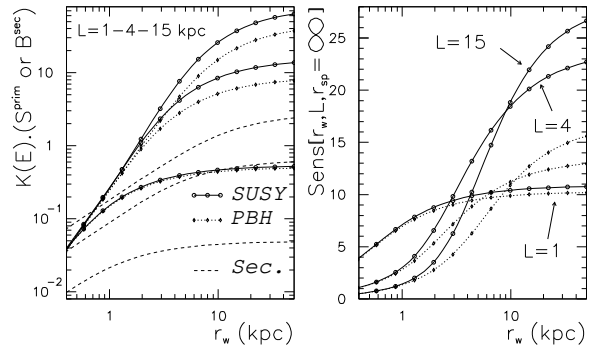}}
\caption{Same quantities as in Fig.~\ref{effet_L}, but as a function
of the propagation parameter $r_{\rm wind}$, and
for three values of $L$ ($r_{\rm sp}=\infty$).}
\label{effet_wind}
\end{figure}
However, it turns out that the effect of the Galactic wind is also
more important for primaries than for secondaries, as the flux is 
exponentially decreased with $z$
for particles created at height $z$ in the diffusive halo.
This can be seen in Fig.~\ref{effet_wind} (left panel).
We clearly see the competition between $L$ and $r_{\rm w}$
in the right panel. For large $L$, the evolution of ${\cal S}ens$ 
is completely driven by $r_{\rm w}$, so that we can compare the result 
to those of Fig.~\ref{effet_L}. When wind is present, 
the sensitivity to a signal
is much more reduced than that we would obtained with a similar $L$
({\em i.e.} a factor $\sim 25$ on ${\cal S}ens$ between the case 
$r_{\rm w}=1$ kpc and $r_{\rm w}=15$ kpc, compared to a factor 3 for $L$
in the same range).
For small $r_{\rm w}$, all primary curves converge to the same value, 
independently of $L$, because then the cosmic rays become blind to 
this boundary, being convected away before having a chance to reach 
the top or bottom of the box.

\subsubsection{The relative rate of spallation through $r_{\rm sp}$} 
\begin{figure}[t]
\centerline{
\includegraphics*[width=\columnwidth]{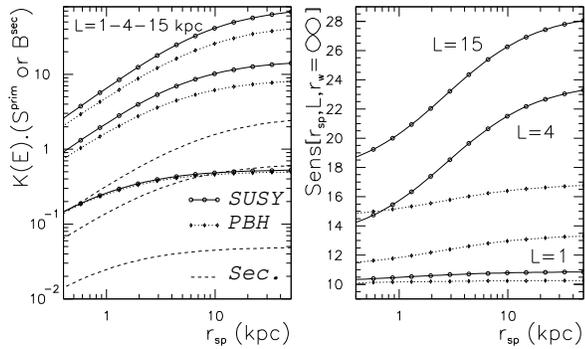}}
\caption{Same quantities as in Fig.~\ref{effet_L}, but as a function
of the propagation parameter $r_{\rm sp}$, and
for three values of $L$ ($r_{\rm w}=\infty$).}
\label{effet_spal}
\end{figure}

At low energy,
particles can be destroyed more easily, because the probability
to cross the disk, and thus to interact with matter, increases 
relatively to the escape (diffusive or convective) probability.
The dependence of ${\cal S}ens$ with $r_{\rm sp}$ is displayed in 
Fig.~\ref{effet_spal}.

When $r_{\rm spal}$ increases, we are sensitive to sources located 
farther, and as for $L$ and $r_{\rm wind}$,
the effect is more important for primaries than for secondaries.
However, the effect of $r_{\rm spal}$ is milder. This is because the 
cut-off due to spallations is less
efficient than escape or convective wind to prevent  particles coming 
from faraway sources from reaching us.

\subsection{A comment about secondaries from GalProp model}
Among several other models that are used to describe cosmic 
ray propagation, the fully numerical approach implemented in 
GalProp \cite{galprop} has been widely used.
Some results obtained within this framework, in particular when studying 
the secondary antiproton spectrum, seem to differ 
(see e.g. Ref. \cite{moska2003}) from ours, obtained with a semi-analytical 
model. 
In our paper we want to derive constraints on the supersymmetric
contribution which can be added to the secondary one, when confronting with
data. Therefore, we take the opportunity of this specific work to 
briefly summarize and discuss some of the differences between the two 
approaches and their results. 

First, the approximation that may appear crucial is that in order
to find analytical expressions for the cosmic ray density, we
have to use a simplified description of the matter distribution in the 
Galaxy, whereas with a numerical approach, any distribution can be considered.
However, the results are not strongly affected by this hypothesis.
In the framework of steady-state diffusion models,
\cite{origin} has shown that the stable nuclei detected
in the solar neighborhood were emitted from sources located in a 
region large enough so that, having sampled very different regions 
of the galactic disc, they are sensitive to a mean density.
Moreover, introducing a radial dependence of the matter distribution
does not induce sizeable difference in the results \cite{riri2003}.
In relation to this first point, we have to emphasize that Ref. 
\cite{moska2003} actually {\em does not} use a detailed description
of the local (i.e. on a scale of a few hundreds of pc) gas distribution.
As a result, they can not provide a reliable analysis of the
radioactive species, which are very sensitive to the local structure 
of the interstellar medium \cite{beta_rad}.

Second, the numerical approach is still costly in terms of computation time,
and is less suited to the systematic study of different effects. 
For example, Ref. \cite{galprop} using a predefined small value $\delta=0.3$ 
for the diffusion coefficient spectral index, finds 
that the observed spectrum of B/C required small values of the Galactic 
wind. Indeed, a full scan of the parameter space, 
extended to a range of values for $\delta$, revealed that models 
with higher values of $\delta$ and with larger values of the galactic 
wind, were actually preferred. This and other results 
 have been thoroughly discussed in Ref. \cite{PaperIb},
and also compared with different propagation models (such as GalProp).
This point is of great importance for the present work, as the theoretical
uncertainties in the antiproton flux are underestimated if some
parameters are not varied over all their plausible values.

The last relevant difference is actually not related to the astrophysical 
model but to the production cross sections. In particular, those relevant 
for B/C have been recently updated (see references in Ref. \cite{moska2003}), 
whereas we use a standard set (see references e.g. in Ref. \cite{PaperI}).
This is a possible way to explain the discrepancy
 between the secondary antiproton flux, but we estimate 
this is unlikely.
Indeed, the two sets of cross sections differ mainly at low energy, 
for which the weight of experimental data is not the greatest.
Using the updated set should not change the propagation parameters 
derived from B/C and used to propagate antiprotons; the final 
results would essentially remain unaffected.

To conclude, we do not see any physically relevant difference between 
the two approaches, and they are probably equally valuable. 
There is still some work to be done from both sides to understand
the origin of the differences in the results, which may lie in the 
methods and interpretation of the results, more than in 
the models themselves.


\section{Results and uncertainties for the primary fluxes}
\label{sec:res_unc}

We now use all the ingredients previously discussed (as well all 
energy changes) to evaluate the primary interstellar  flux.
We try to quantify all the uncertainties that could hamper a
clear selection or exclusion of supersymmetric configurations. They are 
substantially induced 
by the degeneracy in propagation parameters (see 
Sec. \ref{Sec:propagated_flux}) \cite{PaperI,PaperIb} and the choice 
of a peculiar dark matter profile (see Sec. \ref{sec:dm_profile}).

		
\subsection{Primary fluxes and related uncertainties }
\label{Sec:propagated_flux}

The propagation parameters have been constrained by an analysis of the
observed boron over carbon (B/C) ratio, by means of a $\chi^2_{\rm
B/C}$ test over 26 data points and five free parameters
\cite{PaperI}. The best $\chi^2_{\rm B/C}$ was found to be 25.5. A
value of 40 was considered quite conservative, corresponding roughly
to 4-$\sigma$ confidence level on B/C data interpretation, while a
$\chi^2_{\rm B/C}=30$ can be assigned to about 2-$\sigma$ confidence
level \cite{PaperI}.
\begin{figure}[t]
\vspace{-30pt}
{\includegraphics[width=\columnwidth]{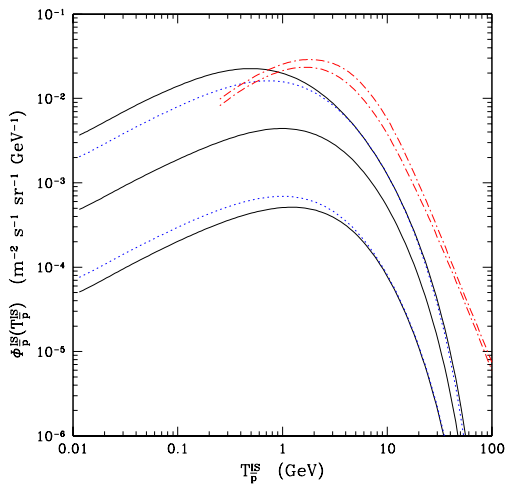}}
\vspace{-30pt}
\caption{The solid lines represent the antiproton flux for a
$m_\chi$=100 GeV neutralino and for  maximal,
median and minimal astrophysical configurations, for
$\chi^2_{\rm B/C}\leq 40$. Dotted lines: the same, but for for
$\chi^2_{\rm B/C}\leq 30$. The dot--dashed band corresponds
to the secondary flux as taken from Ref. \cite{PaperI}
for all the configurations giving $\chi^2_{B/C}\leq 40$.
\label{fig:various_chi2}
}
\end{figure}

In Fig. \ref{fig:various_chi2} we present the result for the primary
antiproton flux for our reference source term for $m_\chi=100$ GeV,
whose $g(E)$ is plotted in Fig. \ref{fig:g_E}. We plot the fluxes corresponding 
to the parameters providing the maximal and minimal fluxes
when all the astrophysical configurations are taken to be compatible
with the analysis on stable nuclei, {\em i.e.}: $\chi^2_{\rm B/C}<40$.  For
{\em the same set} of astrophysical parameters we also plot the
secondary antiproton flux. The variation of the astrophysical
parameters induces a much larger uncertainty on the primary than on the
secondary flux: in the first case, the uncertainty reaches
two orders of magnitude for energies $T_{\bar{p}}\lesssim
1$~GeV, while in the second case it never exceeds 25\% (notice that these
uncertainties are smaller than the nuclear ones, see Ref. \cite{PaperII}).
A thorough discussion about why a combination of parameters gives the
same secondary flux, is skipped here but the reader is referred to
Ref. \cite{PaperIb} for more details. The large variation in the
primary signal can be understood from the previous discussion: first,
the exotic signal is more sensitive to astrophysical parameters than
the standard, as already underlined.  Second, this has to be weighted
by the fact that the secondary flux has in its source term an
additional $K(E)$.  While many combinations of $K_0$, $\delta$, $L$
and $V_c$ lead to the same secondary flux, it is not straigthforward
to decipher which ones lead to the maximum and minimum primary fluxes.
Decreasing $L$ and $r_{\rm w}$ decreases the flux, but at the same
time, to keep the fit to B/C good, $K_0$ has to be also decreased
\cite{PaperI,PaperIb}, increasing in turn the flux (primaries
depends on $1/K(E)$). However, the first two parameters are more important
(especially the wind effect) than the latter. We give in
Tab.~\ref{table:prop} the values for these parameters yielding the
maximum and minimum of the error band in both primary and secondary
fluxes.
\begin{table*}
\[
\begin{array}{|c|c|c|c|c|c|c||c|c|} \hline
{\rm case} &  \delta  & K_0 & L & V_c & V_A & \chi^2_{\rm B/C} 
& r_{\rm w} ({\rm kpc}) & r_{\rm sp} ({\rm kpc}) \\
  & & {\rm (kpc^2/Myr)} & ({\rm kpc}) & ({\rm km/sec}) & ({\rm km/sec}) & &
  	{\rm [1 GeV / 10 GeV]} & {\rm [1 GeV / 10 GeV]}\\\hline \hline
{\rm max} &  0.46  & 0.0765 & 15 & 5   & 117.6 & 39.98 & 29./73. & 
26./57.\\
{\rm med} &  0.70  & 0.0112 & 4 & 12   &  52.9 & 25.68 & 2.4/9.2 & 
4.4/15.\\
{\rm min} &  0.85  & 0.0016 & 1 & 13.5 &  22.4 & 39.02 & 0.33/1.8 & 
0.69/3.1\\
\hline \hline
\end{array} 
\]
\caption{Astrophysical parameters giving the maximal, median and
minimal supersymmetric antiproton flux and compatible wih B/C analysis
($\chi^2_{\rm B/C}<40$).
It is also given in unit of $r_{\rm w}$, $r_{\rm sp}$ (kpc) for two kinetic
energies 1 GeV and 10 GeV.}
\label{table:prop}
\end{table*}
The resulting variation in
Fig. \ref{fig:various_chi2} can be read off from 
Figs.~\ref{effet_L},\ref{effet_wind},\ref{effet_spal},
(left panels) and Tab.~\ref{table:prop}: a factor $\sim 2000$ because of
$r_{\rm w}$ and $L$, an additional factor $\lesssim4$ for $r_{\rm sp}$
(see Fig.~\ref{effet_spal}, left panel) divided by a factor $\sim 50$
because of the value of $K(E)$, leading to an net scattering of
$\sim 100$. This is almost independent on the
specific supersymmetric configuration.

\begin{figure}[t]
\vspace{-30pt}
{\includegraphics[width=\columnwidth]{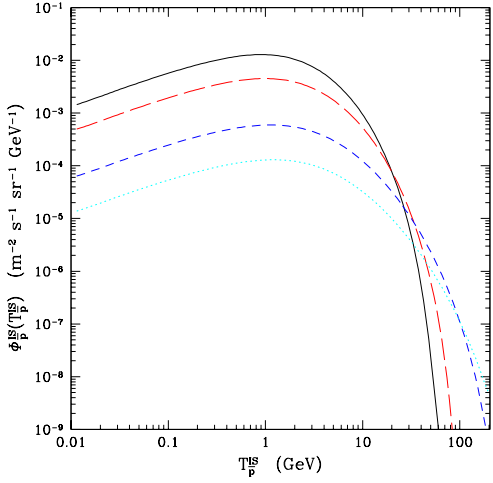}}
\vspace{-30pt}
\caption{Interstellar primary fluxes calculated as a function of the
antiproton kinetic energy. The fluxes are calculated for the median
set of astrophysical parameters. Solid, long dashed, short dashed and
dotted lines correspond to $m_\chi$= 60, 100, 300, 500 GeV,
respectively.  The fluxes correspond to the representative
differential antiproton spectra shown in Fig. \ref{fig:g_E}.
\label{fig:fluxes_med_is}
}
\end{figure}
As emphasized before, energy redistributions relate a specific supersymmetric
configuration (by means of $g(T_{\bar p})$) to the given propagation
configuration. The effect on the resulting antiproton flux is expected
to be mild. We show in Fig. \ref{fig:fluxes_med_is} the result of our
analysis for the representative eMSSM spectra shown in
Fig. \ref{fig:g_E}, corresponding to $m_\chi$= 60, 100, 300, 500 GeV
and for the median astrophysical parameters. The low energy behaviour
of the fluxes is similar for all the masses: this is a consequence of
the propagation of the source spectra, which reduces the intrinsic
differences in the original fluxes at low kinetic energies. The high
energy behaviour of the fluxes reflects the fact that for higher
neutralino masses, the phase space for antiproton production is
larger. Since neutralinos in the Galaxy are highly non--relativistic,
their mass acts as an effective cut--off on the antiproton production
kinetic energy.

\begin{figure}[t]
\vspace{-30pt}
{\includegraphics[width=\columnwidth]{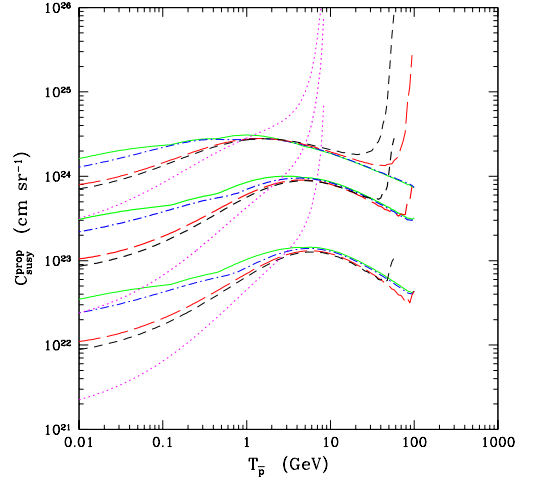}}
\vspace{-30pt}
\caption{Propagation function $C^{\rm prop}_{\rm susy}$ of the primary
supersymmetric antiproton fluxes as a function of the antiproton
kinetic energy, calculated for the reference fluxes of
Figs. \ref{fig:g_E},\ref{fig:fluxes_med_is}. Dotted lines refer to
$m_\chi=10$ GeV, short--dashed to $m_\chi = 60$ GeV, long--dashed to
$m_\chi = 100$ GeV, dot--dashed to $m_\chi = 300$ GeV and solid to
$m_\chi = 500$ GeV. For each set of curves, the upper, medium and
lower line refer to the maximal, median and minimal set of
astrophysical parameters.
\label{fig:C_susy}}
\end{figure}

The effect of propagation on the primary antiproton spectrum may also
be shown by the following function \cite{Bottino_Salati}:
\begin{equation}
C^{\rm prop}_{\rm susy}(T_{\bar{p}})= 
\frac{\Phi_{\bar{p}}(R_\odot,0,T_{\bar{p}})}
{\Upsilon g(T_{\bar{p}})}
\label{eq:C_susy}
\end{equation}
where $\Phi_{\bar{p}}(\odot,T_{\bar{p}})$ is the interstellar
antiproton flux after propagation, normalized to supersymmetric
elementary production term. The propagation function $C^{\rm
prop}_{\rm susy}(T_{\bar{p}})$ is a measure of how the source fluxes
are deformed by propagation and diffusion before reaching the solar
position in the Galaxy and is shown in Fig. \ref{fig:C_susy} for the
same representative spectra of Fig. \ref{fig:g_E}. The energy
dependence is steeper for low--mass neutralinos, and it becomes
somehow more symmetric around a maximal values for neutralinos of
increasing mass. The steep rise of $C^{\rm prop}_{\rm
susy}(T_{\bar{p}})$ near the end of the antiproton production phase
space at $T_{\bar p} = m_\chi$ is due to reacceleration: while the
source factor $g(T_{\bar p})$ is rapidly vanishing, the propagated
flux $\Phi_{\bar{p}}(R_\odot,0,T_{\bar{p}})$ decreases in much milder way
because of reacceleration effects. This effect is more pronounced for
the maximal astrophysical configuration, were $V_A$ is maximal and it
disappears if $V_A$ is set to zero.  Fig. \ref{fig:C_susy} also shows
that the maximal, median and minimal set of astrophysical parameter
affect not only the absolute magnitude of the fluxes but also their
energy dependence: the distorsion of the original flux differs
depending on the values of the propagation parameters, as it has been
discussed in the previous Sections. In particular, the energy of
maximal transfer for neutralino masses above 60 GeV shifts from about
1-2 GeV for the maximal set, to 5-6 GeV for the minimal set.
Fig. \ref{fig:C_susy} shows, at low kinetic energies, a hierarchy in
the behaviour of $C^{\rm prop}_{\rm susy}(T_{\bar{p}})$ which follows
the hierarchy of the neutralino masses: the propagation function is
larger at low enegies for heavier neutralinos, {\em i.e.}  for harder
antiproton fluxes.

The propagation function $C^{\rm prop}_{\rm susy}(T_{\bar{p}})$ can be
directly used to estimate the propagation effects once the supersymmetric
production term $\Upsilon g(T_{\bar{p}})$ is known.



\subsection{Uncertainties related to the dark matter distribution} 
\label{sec:dm_profile}
\begin{table*}[ht]
\begin{center}
\begin{tabular}{|c||c|c|c|}   \hline
   $L$ (kpc), $r_{\rm w}$, $r_{\rm sp}$  & 
\hspace{0.5cm}  
$\displaystyle \frac{S^{prim}_{a=2.5}-S^{prim}_{ref}}{S^{prim}_{ref}}$ 
\hspace{0.5cm} & \hspace{0.5cm} 
  $\displaystyle \frac{S^{prim}_{a=5}-S^{prim}_{ref}}{S^{prim}_{ref}}$ 
\hspace{0.5cm} &\hspace{0.5cm} 
  $\displaystyle \frac{S^{prim}_{NFW}-S^{prim}_{ref}}{S^{prim}_{ref}}$ 
\hspace{0.5cm}
\\\hline\hline
15, 28.66, 25.54 &
 -69.5\%  & +23.9\% & +19\%\\\hline
4, 2.38, 4.41
	& -21.5\% & +9.9\%  & $\sim 0\%$ \\\hline
1, 0.33, 0.69 
	& $<1\%$ & $<0.2\%$    &  $\sim 0\%$ \\\hline
\end{tabular}
\caption{Sensitivity to the core radius of an isothermal profile, and
comparison of the NFW and istothermal profiles, for three
representative propagation sets at $T_{\bar p}= 1$ GeV. These
propagation parameters correspond to the minimum, median and maximum
primary flux compatible with nuclei analysis. The reference value
$S^{prim}_{ref}$ is for an isothermal halo whose core radius is
$a=3.5$ kpc.  Notice that for higher energies, the results would be
the same than those provided by the set $L=15$~kpc (purely diffusive
transport).}
\label{tab:core}
\end{center}
\end{table*}

We have performed all the calculations assuming that the Galactic dark
matter is distributed as an isothermal sphere with a core radius
$a=3.5$~kpc and local dark matter density $\rho_l=0.3$ GeV~cm$^{-3}$.
For this density profile, we have estimated that the antiproton
propagation induces an uncertainty on the primary antiproton flux of
about two orders of magnitude, especially at low kinetic energies.

Another source of uncertainty on the primary flux comes from the shape
of the dark matter density profile, which is only poorly known, and
from the allowed range of values of $\rho_l$ for any given density
distribution. We have already commented that for an isothermal
spherical distribution, the local dark matter density may range from
0.18~GeV~cm$^{-3}$ to 0.71~GeV~cm$^{-3}$. Moreover, the dark matter
distribution may be quite different from a simple isothermal sphere
(see, for instance, Refs. \cite{moore,nfw,dehnen,olling,bcfs} and
references therein): the cold dark matter distribution could be
non--spherically symmetric, it can be singular at the Galactic center,
as suggested from numerical simulations, or it can present a clumpy
distribution in addition to a smooth component. Since the shape of the
Galactic halo enters as $\rho^2_{\rm DM}(r,z)$ in the evaluation of
the astrophysical part for the primary signal $S^{\rm prim}_{\rm
astro} (R_{\odot},0,E)$, it is a main ingredient in the determination
of the primary antiproton flux, and the uncertainties in the
description of $\rho_{\rm DM}(r,z)$ may sizeably affect the predicted
signal.

The uncertainty in $\rho_l$, determined by a detailed modelling of the
Galactic component \cite{dehnen,olling} and mainly due to the value of
the local rotational velocity \cite{bcfs}, depends on the shape of the
Galactic halo. For the same isothermal sphere, the range
in $\rho_l$ may change the primary fluxes by a factor which ranges
from 0.36 to 5.6: overall, even for the simple choice of an isothermal
sphere, the antiproton flux has an uncertainty of a factor of about
15, on top of the two orders of magnitude due to antiproton
propagation. We anticipate that, among all the uncertainties due to
the shape of the Galactic halo, the uncertainty coming from $\rho_l$
will turn out to be the most relevant one (apart, eventually, the
presence of close clumps).

Independently of the normalization $\rho_l$, any given density profile
could, in principle, modify the signal. In particular, distribution
functions derived from numerical simulations are singular towards the
Galactic center, where a very high neutralino annihilation rate would
then occur. We could thus expect that such dark matter profiles would
induce an enhanced antiproton flux with respect to a non--singular
distribution. In this class of modified density profiles we can also
include an isothermal sphere with different values of the core radius
$a$. We expect that enlarging the core radius would increase the
signal. We therefore estimated the modification of the cosmic
antiproton flux when different core radii and dark matter profiles are
used in the source term. The reference flux is obtained with our
spherical isothermal distribution, with core radius $a=3.5$ kpc. The
results are shown in Table \ref{tab:core}. Notice that we have used
for all the profiles the normalization $\rho_l=0.3$ GeV~cm$^{-3}$, in
order to extract the change on the antiproton flux which is due
entirely to the different shapes of the halos. It is clear that each
density profile will have to be further implemented with its own value
of $\rho_l$ \cite{dehnen,olling,bcfs}. From Table \ref{tab:core} we
notice, first of all, that for small $L$ and $r_{\rm w}$, we are
completely blind to what occurs near the Galactic center. Only the
very local properties of the dark matter distribution are of some
relevance for our study. For a diffusive halo of 4 kpc, we varied the
core radius of the isothermal distribution from 2.5 to 5 kpc. With
respect to our reference values of 3.5 kpc, small $a$ leads to a
reduction of the flux by about $20\%$, while large values of $a$ give
a 10\% increase. For $L=15$ kpc -- and all the other propagation
parameters modified consequently -- a 2.5 kpc core radius diminishes
the reference flux by 70\% and a 5 kpc one pushes it up by 25\%.  The
uncertainty of a factor two on the core radius of the isothermal
distribution then reflects in a factor of four indeterminacy of the
primary antiproton flux. As for the singular density profiles, Table
\ref{tab:core} shows that a NFW \cite{nfw} distribution function does not
strongly modify the flux: when compared to the isothermal case, the
flux is increased by no more than about 20\%, and this occurs when the
diffusive halo size is the largest.  For $L\lsim$5 kpc, the difference
between an isothermal profile and a NFW singular distribution is
irrelevant. This result clearly shows that it is very improbable for
an antiproton produced at the Galactic center to reach the Earth.

Finally, one can deal with halos which contain regions of enhanced
density called clumps. In this sub--halos, the neutralino annihilation
is more effective and the signal can be increased by some enhancement
factor. However, as also suggested by Ref. \cite{berezinsky}, this
enhancement is not propagation dependent and simply acts on the
antiproton flux as a normalization factor. From the analysis of
Ref. \cite{berezinsky}, the average enhancement is likely to be
smaller than a factor of 5. A detailed analysis of this point is
beyond the scope of our paper; however, the effects of such an
enhancement are briefly discussed at the end of Sec.
\ref{sec:toa_susy}.

In conclusion, we wish to remark that our choice of an isothermal
sphere with a core radius $a=3.5$~kpc and local dark matter density
$\rho_l=0.3$ GeV~cm$^{-3}$, together with the best choice for the
astrophysical parameters which govern diffusion and propagation in the
Galaxy, represent an optimal choice for the prediction of the
antiproton signal. Our results will not be dramatically modified by a
different choice for the density profile, while a different choice for
the local dark matter density is easily taken into account as a
normalization factor.

%

%
%
\section{Top--of--atmosphere fluxes: comparison with data and results for 
supersymmetric models}
\label{sec:toa_susy}

\begin{figure}[t]
\vspace{-30pt}
{\includegraphics[width=\columnwidth]{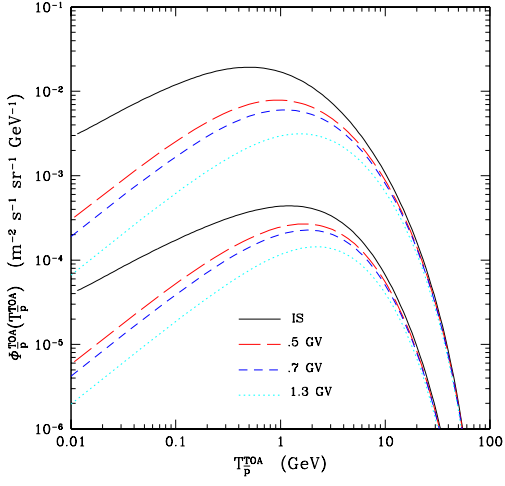}}
\vspace{-30pt}
\caption{Top--of--atmosphere antiproton fluxes as a function of the
antiproton kinetic energy for the $m_\chi$=100 GeV reference case.
The upper (lower) set of curves refer to the maximal (minimal) set of
of astrophysical parameters. Solid curves show the interstellar
fluxes. Broken curves show the effect of solar modulation at
different periods of solar activity: $\phi=500$ MV (long dashed),
$\phi=700$ MV (short dashed), $\phi=1300$ MV (dotted).
\label{fig:flux_solmod}
}
\end{figure}
\begin{figure}[t]
\vspace{-30pt}
{\includegraphics[width=\columnwidth]{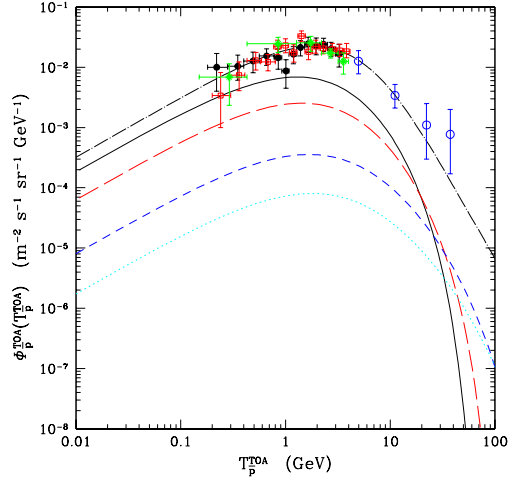}}
\vspace{-30pt}
\caption{Primary TOA antiproton fluxes as a function of the antiproton
kinetic energy, for the representative spectra of Figs. \ref{fig:g_E}
in the eMSSM. The solid line refers to $m_\chi = 60$ GeV, the
long--dashed line to $m_\chi = 100$ GeV, the short--dashed line to
$m_\chi = 300$ GeV and the dotted line to $m_\chi = 500$ GeV. The
astrophysical parameters correspond to the median choice.  Solar
modulation is calculated for a period of minimal solar
activity. The upper dot--dashed curve corresponds to the antiproton
secondary flux taken from Refs. \cite{PaperII,revue}. Full circles show the
{\sc bess} 1995-97 data \cite{bess95-97}; the open squares show the {\sc bess}
1998 data \cite{bess98}; the stars show the {\sc ams} data \cite{ams98} and
the empty circles show the {\sc caprice} data \cite{caprice}.
\label{fig:prim_sec_data_med_solmin}
}
\end{figure}

\begin{figure}[t]
\vspace{-30pt}
{\includegraphics[width=\columnwidth]{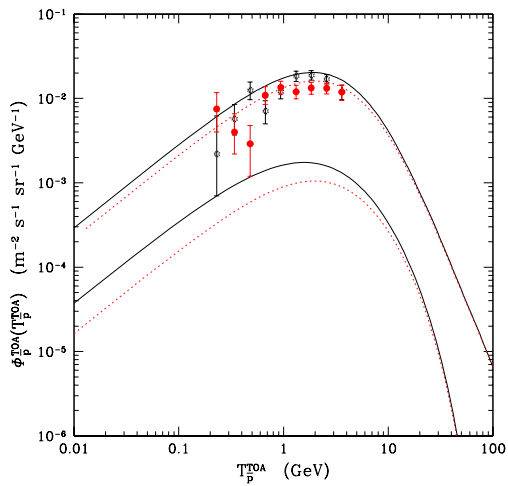}}
\vspace{-30pt}
\caption{Primary TOA antiproton fluxes at solar maximum for the
transient periods of solar activity of years 1999 and 2000. The upper
set of curves show the antiproton secondary fluxes. The lower set of
curves show the primary antiproton fluxes obtained for the
representative $m_\chi=100$ GeV case. The solar modulation parameter
is fixed at 700 MV (solid lines) and at 1300 MV (dotted lines).  The
astrophysical parameters correspond to the median case.
Stars and full circles correspond to {\sc bess} 1999 and 2000, respectively
\cite{bess99-00}.
\label{fig:prim_sec_data_solmax}
}
\end{figure}
\begin{figure}[t]
{\includegraphics[width=\columnwidth]{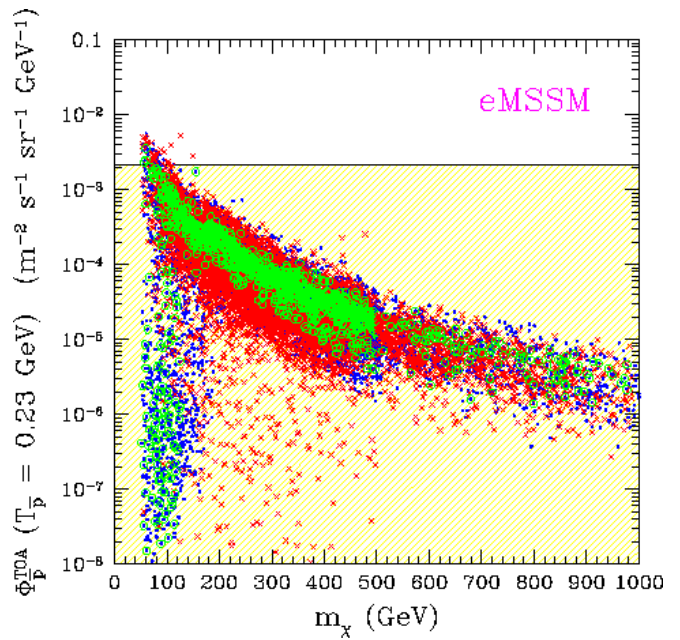}}
\caption{\label{fig:mssm_pbar023}Antiproton flux at solar minimum from
neutralino annihilation calculated at $T_{\bar p}=0.23$ GeV, as a
function of the neutralino mass for a generic scan of the eMSSM. The
flux is calculated for a smooth halo described by an isothermal
profile with core radius $a=3.5$ kpc and for the median set of
astrophysical parameters.  Crosses (in red) refer to cosmologically
dominant neutralinos ($0.05 \leq \relic \leq 0.3$); dots (in blue)
refer to subdominant relic neutralinos ($\relic < 0.05$); light
circles (in green) show the eMSSM configurations for which the
neutralino relic abundance lies in the preferred range for CDM, as
determined by the combined {\sc wmap}+2d{\sc fgrs}+Lyman--$\alpha$ analysis:
$0.095 \leq \Omega_{CDM} h^2 \leq 0.131$ \cite{cmb}. The shaded region
(in yellow) denotes the amount of antiprotons, in excess of the
secondary component \cite{PaperII}, which can be accomodated at
$T_{\bar p}=0.23$ GeV in order not to exceed the observed flux, as
measured by {\sc bess} \cite{bess95-97,bess98}. All the points of the
scatter plot which lie below the horizontal black line are compatible
with observations.}
\end{figure}
\begin{figure}[t]
{\includegraphics[width=\columnwidth]{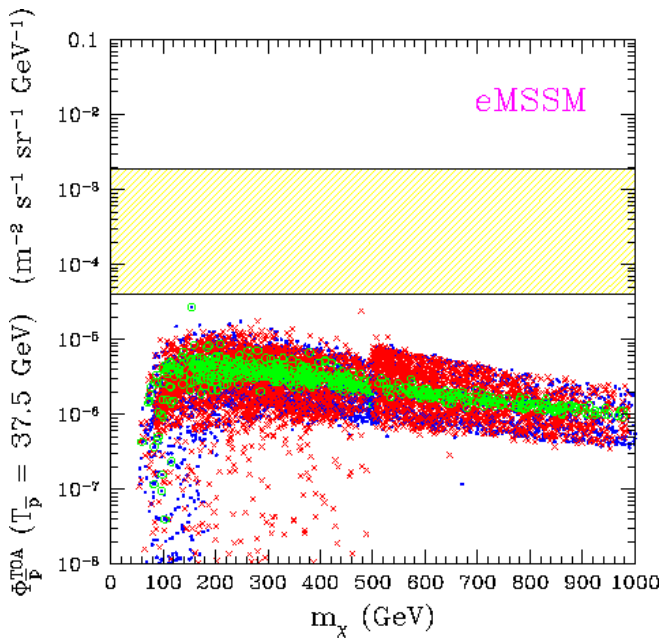}}
\caption{\label{fig:mssm_pbar375} Antiproton flux at solar minimum
from neutralino annihilation calculated at $T_{\bar p}=37.5$ GeV, as a
function of the neutralino mass for a generic scan of the
eMSSM. Notations are as in Fig. \ref{fig:mssm_pbar023}. The shaded
region (in yellow) denotes the amount of antiprotons which would be
required at $T_{\bar p}=37.5$ GeV in order to explain the possible
excess in the {\sc bess} data \cite{bess95-97,bess98} over the secondary
component \cite{PaperII}. All the points of the scatter plot which lie
below the upper horizonthal black line are compatible with
observations.}
\end{figure}
\begin{figure}[t]
{\includegraphics[width=\columnwidth]{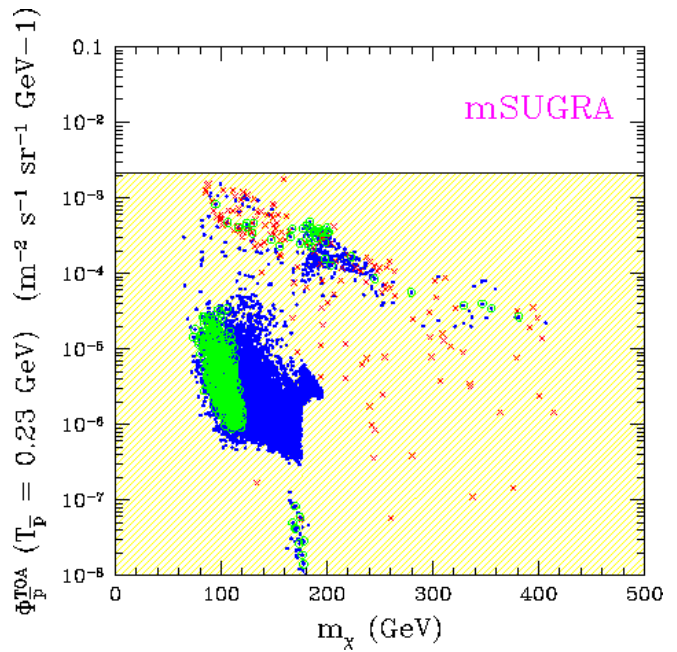}}
\caption{\label{fig:sugra_pbar023}
The same as in Fig. \ref{fig:mssm_pbar023}, for
 a generic scan of the mSUGRA scheme.}
\end{figure}
\begin{figure*}
{\includegraphics[width=\textwidth]{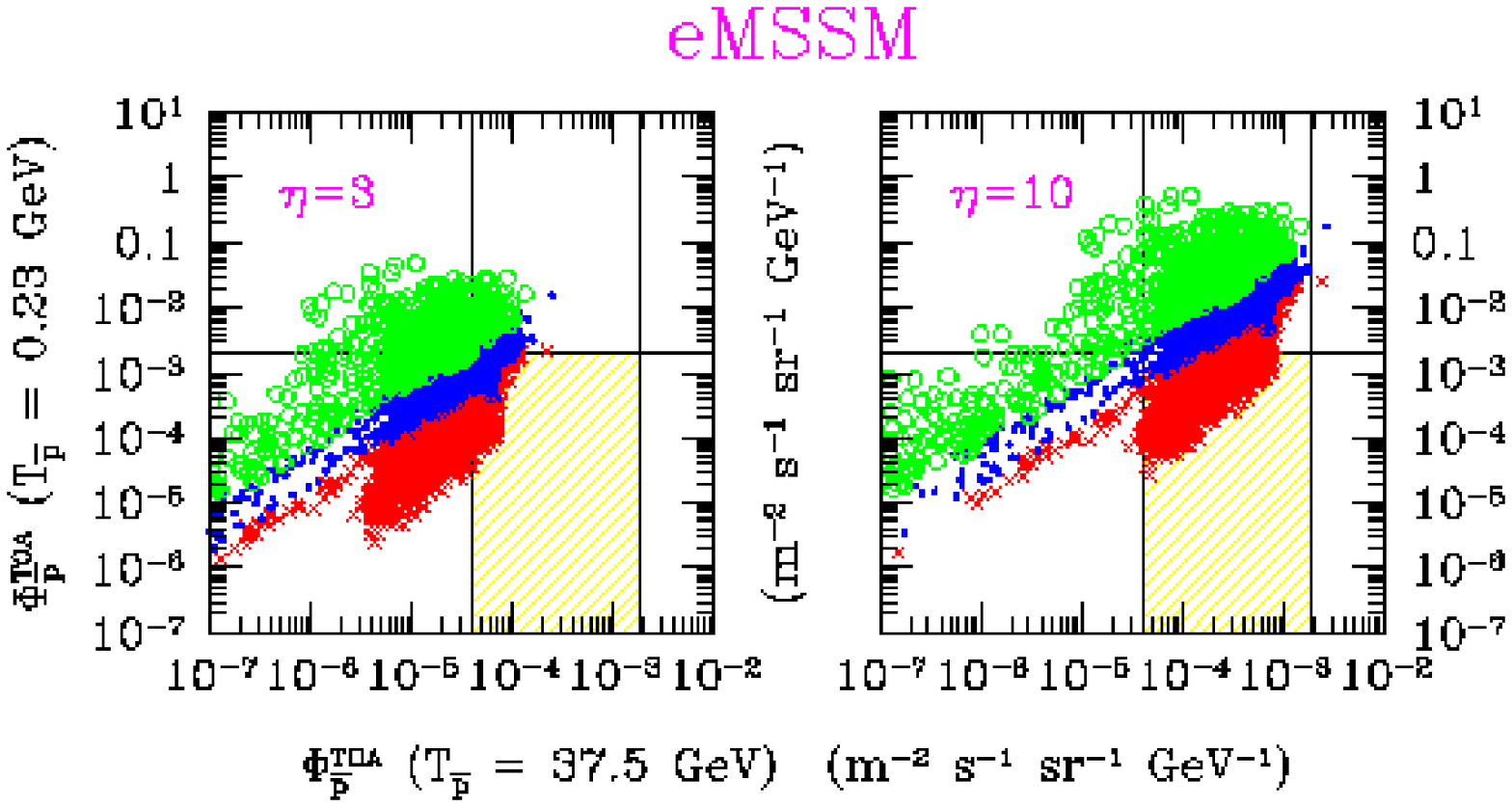}}
\caption{\label{fig:mssm_fill}Correlation between the antiproton flux
at $T_{\bar p}=0.23$ GeV and $T_{\bar p}=37.5$ GeV shown in
Figs. \ref{fig:mssm_pbar023} and \ref{fig:mssm_pbar375} for the
astrophysical enhancement parameter $\eta=3$ and 10 (overdense halos)
and for the median set of astrophysical parameters.  Circles (in
green), dots (in blue) and crosses (in red) denote configurations with
neutralino masses in the ranges: 50 GeV $< m_\chi <$ 150 GeV, 150 GeV
$<m_\chi <$ 300 GeV and 300 GeV $< m_\chi <$ 1 TeV, respectively.  The
horizonthal line denotes the upper limit on the antiproton flux at
$T_{\bar p}=0.23$ GeV coming from the {\sc bess} data
\cite{bess95-97,bess98}, once the secondary component \cite{PaperII}
is taken into account. The rightmost vertical line denotes the
corresponding upper limit at $T_{\bar p}=37.5$ GeV.  The shaded area
indicates configurations which can explain the possible excess in the
data \cite{bess95-97,bess98} over the secondary
component \cite{PaperII} at $T_{\bar p}=37.5$ GeV, without giving an
excees at low kinetic energies ($T_{\bar p}=0.23$ GeV).}
\end{figure*}

Now that we have calculated the interstellar fluxes of antiprotons at
the Sun's position in the Galaxy, we have to further propagate them
inside the heliosphere, where the cosmic ray particles which
eventually reach the Earth are affected by the presence of the solar
wind. We model the effect of solar modulation by adopting the force
field approximation of the full transport equation
\cite{Perko}. In this model, the top--of--atmosphere (TOA) antiproton
flux $\Phi^{\rm TOA}_{\bar{p}}$ is obtained as:
\begin{equation}
\frac{\Phi^{\rm TOA}_{\bar p} (E^{\rm TOA}_{\bar p})}{\Phi^{\rm IS}_{\bar p} (E^{\rm IS}_{\bar p})}
= 
\left( \frac{p^{\rm TOA}}{p^{\rm IS}} \right)^{2}\;
\end{equation}
where $E$ and $p$ denote the total energies and momenta of
interstellar and TOA antiprotons, which are related by the energy shift:
\begin{equation}
E^{\rm IS}_{\bar p} = E^{\rm IS}_{\bar p} - \phi
\end{equation}
where the parameter $\phi$ is determined by fits on cosmic ray
data. In our analysis, we will adopt the value $\phi=500$ MV for
periods of minimal solar activity, corresponding to the years around
1995--1998, $\phi=700$ and $\phi=1300$ MV for a transient period and
for the solar maximum, respectively, which will be used for years
1999 and 2000.

Fig. \ref{fig:flux_solmod} shows the TOA antiproton fluxes for the
$m_\chi=100$ GeV reference configuration and for the maximal and
minimal sets of astrophysical parameters. The figure shows that solar
modulation has the effect of depleting the low--energy tail of the
antiproton flux. The effect is clearly more pronounced for periods of
stong solar activity, when the solar wind is stronger.

Data on antiprotons at Earth are now abundant, mostly after the
missions of the balloon borne detector {\sc bess}. This experiment has
provided measurements at different periods of the solar activity
\cite{bess95-97,bess98,bess99-00}. It has now collected more than two
thousand antiprotons between 200 MeV and 4 GeV.  Data at solar minimum
have been taken also by the {\sc ams} experiment on board of Shuttle
\cite{ams98} in an energy range similar to {\sc bess}, and by the 
{\sc caprice} balloon at higher energies, namely between 5 GeV and 40 GeV
\cite{caprice}. All the data at solar minimum are plotted in Fig.
\ref{fig:prim_sec_data_med_solmin} along with the secondary reference
flux (for details, see Ref. \cite{PaperII}) and our predictions for
primary fluxes at different neutralino masses in the eMSSM: $m_\chi =
60,100,300,500$ GeV and for the median set of astrophysical
parameters. We notice that the primary flux from neutralino
annihilation is at most of the same order of magnitude as the
secondary flux, and this occurs for neutralino masses close to their
current lower bound in the eMSSM, which is around $m_\chi \simeq 50$
GeV. We remind that the representative supersymmetric configurations
plotted in Fig. \ref{fig:prim_sec_data_med_solmin} refer to a large
antiproton production for each mass ({\em i.e.} they correspond to
large values of the $\Upsilon$ parameter shown in
Fig. \ref{fig:mssm_y_mchi}). This indicates that the antiproton signal
for neutralino dark matter will hardly produce an excess over the
secondary flux, for an isothermal matter profile of the Galactic halo
and for $\rho_l=0.3$ GeV cm$^{-3}$. This occurs for the median (and
best) choice of the astrophysical parameters which govern the
diffusion and propagation of antiprotons in the Galaxy. Clearly, the
maximal set of astrophysical parameters, which produces fluxes about
one order of magnitude larger than the median set, may produce a large
excess, for neutralino masses below 100-200 GeV. This excess could
then be used to constrain supersymmetric models since the secondary
flux is perfectly compatible with the data. However, for setting
constraints on supersymmetry in a conservative way, we should instead
use the set of astrophysical parameters which produces the minimal
fluxes. In this case, the primary fluxes are lower than the ones
plotted in Fig. \ref{fig:prim_sec_data_med_solmin} by about one order
of magnitude, as discussed in the previous Section.  In conclusion,
our analysis shows that, due to the large uncertainties in the primary
fluxes, the antiproton signal is not suitable at present for setting
{\em conservative} constraints on supersymmetric models. For this we
need a better knowledge of the astrophysical parameters that govern
the diffusion and propagation of primary antiprotons in the Galaxy.

Antiproton data are also available for periods of intense solar
activity from the {\sc bess} detector.  Fig.
\ref{fig:prim_sec_data_solmax} shows these data together with the
secondary flux and the primary flux calculated for the representative
$m_\chi=100$ GeV configuration.  The astrophysical parameters are
fixed at their median values. Also at solar maximum, we see that the
secondary flux is compatible with the data and the supersymmetric flux
is sensitively smaller than the secondary one.

In our discussion so far, we have commented that {\em conservative}
and solid constraints on supersymmetric models require the use of the
minimal set of astrophysical parameters.  This attitude is needed in
setting limits.  However, the best and most probable choice of
astrophysical parameters is the median one, and we will therefore
adopt from here on this set of parameters for our analyses. It is
likely that a sharpening of the knowledge of the propagation
parameters will lead to a shrinking of the allowed uncertainty band
around the central (median) value.

In order to compare the experimental results with a full scan of the
supersymmetric parameter space, we calculate the TOA antiproton fluxes
in two different energy bins and compare our results with the excess
which can be accomodated above the secondary flux in order not to
enter in conflict with the experimental data in that energy bin.  We
have chosen a low energy bin: $T_{\bar p} = 0.23$ GeV, and a high
energy one: $T_{\bar p} = 37.5$ GeV. As can be seen in
Fig. \ref{fig:prim_sec_data_med_solmin}, in the low energy bin the
secondary flux is perfectly compatible with the data, therefore no
excess is needed: this allows us to set an upper bound on the possible
amount of antiprotons of primary origin which can be accomodated:
$\Phi^{\rm TOA}_{\bar p}(T_{\bar p} = 0.23 {\rm ~GeV}) \lsim 2.09\cdot
10^{-3}$ m$^{-2}$ s$^{-1}$ sr$^{-1}$ GeV$^{-1}$.  This value is
obtained by taking into account the values and uncertainties of both
data and secondary flux at $T_{\bar p} = 0.23$ GeV. At $T_{\bar p} =
37.5$ GeV, even though the data and the secondary flux are
statistically compatible, a possible excess may be accomodated, since
the central value of the experimental point indicates a much larger
flux as compared to the secondary component. In this case, we can
define an interval of values for a possible excess: $0.04 \cdot
10^{-3} \lsim \Phi^{\rm TOA}_{\bar p}(T_{\bar p} = 37.5 {\rm ~GeV})
\lsim 1.87\cdot 10^{-3}$ in units of: m$^{-2}$ s$^{-1}$ sr$^{-1}$
GeV$^{-1}$. We compare these intervals with our calculations in the
eMSSM and in mSUGRA.

 Fig. \ref{fig:mssm_pbar023}
shows the scatter plot of the antiproton
flux calculated at $T_{\bar p} = 0.23$ GeV for a generic scan of the
eMSSM scheme. The supersymmetric fluxes are clearly largest at low
neutralino masses, and they fall down as the neutralino mass increases
mostly because the neutralino number density in the Galaxy scales as
$m_\chi^{-2}$. A small fraction of configurations with masses below
100 GeV can provide fluxes which could be potentially at the edge of
producing an excess, but we remind here that for a safe exclusion
of these configurations we should use the minimal set of astrophysical
parameters, which provides a flux which is about one order of magnitude
smaller. In any case, a reduction of the uncertainties on the primary
flux calculation and a future reduction of experimental errors may
eventually either allow to set limits to supersymmetry or show a
positive excess of antiprotons in this low--energy bin, a fact which
could then be explained as originated by neutralinos of masses below 100
GeV. 
Fig. \ref{fig:mssm_pbar375} shows the scatter plot of the
antiproton flux calculated at $T_{\bar p} = 37.5$ GeV for the same
scan of the eMSSM. In this case we observe that all the supersymmetric
configurations are compatible with data, but there are no
supersymmetric models which can allow us to explain the discrepancy
between the data and the secondary flux as due to an excess of
supersymmetric origin.

The situation in the mSUGRA scheme is shown in
Fig. \ref{fig:sugra_pbar023} with the flux of antiprotons at
$T_{\bar p} = 0.23$ GeV. In this case, as already observed in
connection with the properties of the mSUGRA source term, the
antiproton fluxes are smaller than in the eMSSM case. Nevertheless, a
restricted fraction of mSUGRA configuration is potentially explorable
in the future, with a reduction of the experimental error of about a
factor of 2-3.

The fluxes we have shown so far, all refer to a dark matter density
distribution in the form of an cored isothermal sphere. Clearly a halo
profile which is able to produce an overdensity with respect to the
isothermal sphere would produce a larger antiproton flux. 
We can parameterize the enhancement of the matter density by a
multiplicative factor $\eta$, which then enters as $\eta^2$ in the
calculation of the antiproton primary flux, since the flux depends on
the square of the matter density. The origin of the overdensity may be
due, for instance, to flattening of the Galactic halo or to the
presence of clumps. It the latter case, the enhancement factor is
likely to be smaller than about 5, once the result of
Ref. \cite{berezinsky} are implemented with our discussion on the
antiproton diffusion region in the Galaxy. The enhancement factor may
also be related to a different choice of the local dark matter
density, which has been fixed at $\rho_l=0.3$ GeV~cm $^{-3}$ in our
analysis. We remind that our cored isothermal sphere allows factors of
$\eta$ up to: $\eta \sim (0.71/0.3) = 2.4$ \cite{bcfs}.  Clearly, a
complete reanalysis of the propagation and diffusion properties will
be required for each different choice of the halo shape: this
reanalysis will give the amount of enhancement concerning the specific
halo.  Anyhow, regardless of how the enhancement $\eta$ is obtained,
we can discuss the effect of such an increased flux by using $\eta$ as
a normalization factor to show the amount of enhancement which is
required in order to interpret the antiproton excess at $T_{\bar
p}=37.5$ GeV as due to neutralino dark matter, without exceeding the
upper limit on the antiproton flux at $T_{\bar p}=0.23$ GeV.
Fig. \ref{fig:mssm_fill} shows the correlation between the eMSSM
antiproton fluxes at $T_{\bar p}=0.23$ GeV and $T_{\bar p}=37.5$ GeV
for  $\eta$=3 and 10. 
The supersymmetric configurations which could fulfill this requirement are
the ones which fall inside the shaded area in Fig. \ref{fig:mssm_fill}. 
We see that the possible excess at $T_{\bar
p}=37.5$ GeV requires halo overdensities of at least a factor of 2-3
and neutralino masses larger than a few hundreds of GeV. This last
property is simply understood on the basis of the fluxes shown in
Fig. \ref{fig:prim_sec_data_med_solmin}: the phase space cut off at
$T_{\bar p} = m_\chi$ implies that light neutralino would need a huge
overdensity factor in order to match the observed antiproton flux at
$T_{\bar p}=37.5$ GeV, but this would produce an exceedingly large
flux at $T_{\bar p}=0.23$ GeV. On the contrary, heavy neutralinos have
a phase space cut off at much higher kinetic energies, and therefore a
mild overdensity may enhance the flux at $T_{\bar p}=37.5$ GeV without
giving conflict at low kinetic energies.

\section{Conclusions and perspectives}
\label{sec:conclusion}

We have calculated the flux of antiprotons produced by relic
neutralino annihilations in the Galactic halo in a detailed diffusion
model constrained by analysis of stable and radioactive nuclei. The
source of antiprotons is studied both in a low--energy minimal
supersymmetric standard model (eMSSM) and in a supergravity inspired
supersymmetric scheme (mSUGRA).  We find that the interstellar primary
antiproton fluxes are affected by a large uncertainty, which spans two
orders of magnitude at low antiproton kinetic energies. This is at
variance with the secondary antiproton fluxes (whose uncertainty never
exceeds 24\% \cite{PaperI}) and it is mainly related to the fact that
the source of the primary flux is located inside the diffusive halo,
whose size is unknown. By adopting a conservative choice for the dark
matter density--profile and propagation parameters, no supersymmetric
configuration can be excluded on the basis of an excess over the
existing data. Actually, the data are  quite well explained by
the secondary contribution alone. However, if we adopt the best values
for the propagation parameters (corresponding to a thickness of the
diffusive halo of 4 kpc), a window of low--mass neutralino
configurations provides fluxes which, once summed up to the secondary
contribution, are in excess of the experimental measurements. We have
shown that the sensitivity to the antiproton signal is increased with
the halo size and limited by strong convection. An improved knowledge
of the propagation parameters will certainly help in reducing the
uncertainty on the primary flux and consequently it could allow us to
set more severe constraints on the supersymmetric parameter space.

The sensititity of the primary antiproton flux on the shape of the
dark matter density profile has also been investigated. We have shown
that the shape of the dark matter density distribution does not
introduce large uncertainties. In particular, we have demonstrated
that a NFW distribution can increase the primary antiproton flux by no
more than 20\% with respect to the isothermal profile.  Indeed, it is
very improbable to detect at Earth antiprotons produced in the
centeral regions of the Galaxy, where the two distributions differ
most.

In the next years several balloon-borne experiments such as {\sc
bess}, space--based detectors such as {\sc ams} and satellites as {\sc
pamela} will provide very abundant and accurate data on the antiproton
flux.  At the same time, new data on cosmic ray nuclei are expected
and would lead to a better knowledge of the cosmic--ray propagation
mechanisms. We could thus expect a dramatic reduction of the
uncertainties affecting the neutralino-induced antiproton flux and
much more definite predictions for antiprotons of supersymmetric origin
will then be possible. 
Many efforts are also devoted to other indirect neutralino searches, such 
as positrons and antideuterons in cosmic rays, gamma rays and up-going 
muons, as well as to direct searches in deep underground laboratories, 
giving thus the hope that more constraining analysis on the existence 
of relic neutralinos in the halo of our Galaxy will be viable.



\appendix

\section{Calculation of the antiproton differential spectrum per
annihilation event}
\label{sec:appendix}

The antiproton differential spectrum per annihilation event $g(T_{\bar
p}$) is calculated by following analytically the decay chain of the
neutralino annihilation products until a quark or a gluon $h$ is
produced. The antiproton spectrum is then obtained by a Monte Carlo
modelling of the quark and gluon hadronization (we make use of the
PYTHIA package \cite{pythia}). We have produced the $\bar p$
differential distributions for $h=u,d,s,c,b,{\rm gluon}$ at various
injection energies for each $h$ (the $t$ quark is assumed to decay
before hadronization and is treated analytically, since in addition to
its standard model decay into $W^+b$, it may have a supersymmetric
decay into $H^+b$). Whenever we need the $\bar p$ distribution for an
injection energy different from the produced ones, we perform an
interpolation. In order to obtain the antiproton differential
distribution in the neutralino rest frame we perform the necessary
boosts on the MC spectra.

For instance, let us consider a ${\bar p}$ production from a neutralino
decay chain of this type:
\begin{equation}
\chi\chi \rightarrow A \rightarrow a \rightarrow h \rightsquigarrow {\bar
p}\;\;.
\label{eq:decay}
\end{equation}
The antiproton differential spectrum per annihilation event $g(T_{\bar
p}$) is then obtained by the product of the branching ratios for the
production of $A$, $a$ and $h$ in the decay chain, with the
differential distribution of antiprotons produced by the hadronization
of an $h$ injected at an energy $E_{\rm prod}$ (defined in the rest
frame of the $a$ decaying particle) double boosted to the $\chi$
reference frame:
\begin{eqnarray}
g(T_{\bar p}) &=& 
{\rm BR}(\chi\chi \rightarrow A){\rm BR}(A \rightarrow a){\rm BR}(a \rightarrow h)\times \nonumber \\
&& \left[\left({dN^h_{\bar p}\over dT_{\bar p}}
        \right)_{{\rm boost~}a\rightarrow A}
   \right]_{{\rm boost~}A\rightarrow\chi}\;\;.
\end{eqnarray}
The first boost transforms the spectrum from the rest frame of $a$ (in
which $h$ is injected with energy $E_{\rm prod}$) to the rest frame of
$A$. The second brings the distribution to the rest frame of
$\chi$. Each boost is obtained by the following expression:
\begin{equation}
g(E_{\bar p}) = \frac{1}{2} \int_{E'_-} ^{E'_+} 
\left. \left(\frac{dN^h_{\bar p}}{dE'}\right) \right |_{E_{\rm prod}}
\, \frac{dE'}{\gamma \beta\ p'}
\end{equation}
where $E_{\bar p} = T_{\bar p}+m_{\bar p}$ is the total antiproton
energy, $p'=\sqrt{(E'^2+m_{\bar p}^2)}$, $\gamma$ and $\beta$ are the
Lorentz factors of the boost and the interval of integration is
defined by:
\begin{equation}
E'_{\pm} = \min\left[
E_{\rm prod}, \gamma E_{\bar p}
\left( 1\pm \beta\sqrt{1-\frac{m_{\bar p}^2}{E_{\bar p}^2}}
\right)
\right]\;\;.
\end{equation}
%


\section{Solution of the diffusion equation}
\label{sec:appendix2}

In cylindrical geometry, the differential 
density $N^{\bar{p}}(r,z,E)$ is given by
\begin{eqnarray}
     0 &=&
     \left[K(E)\left(\frac{\partial^{2}}{\partial z^{2}} +
     \frac{1}{r}\frac{\partial}{\partial r}
     \left(r\frac{\partial}{\partial r} \right) \right)
     - V_{c} \frac{\partial}{\partial z}\right] N^{\bar{p}}(r,z,E)  \nonumber \\
     &+& {\cal Q}^{\bar{p}}(r,z,E)
     - 2h\,\delta(z)\, \Gamma^{\bar{p}}(E)\,N^{\bar{p}}(r,z,E)\;\;,
     \label{transport_CR}
\end{eqnarray}
where the energy losses have been omitted, for the sake of clarity.
The source term includes primary antiprotons -- 
from exotic sources present in the 
dark halo, annihilating throughout the diffusive halo of the Galaxy --, 
and secondary antiprotons -- standard p and He CRs spallating on the 
interstellar gas in the thin disk --, and may be written as
\begin{equation}
     \label{SOURCES}
     {\cal Q}^{\bar{p}}(r,z,E) = q^{\bar{p},{\rm prim}}(r,z,E)+
     2h\delta(z)q^{\bar{p},{\rm sec}}(r,0,E)\;\;.
\end{equation}

A convenient way to solve Eq.~(\ref{transport_CR}) is to expand all the 
functions $f(r)$ (the density
$N(r)$ and the source distribution $q(r)$) that depend on $r$ on
the orthogonal set of  Bessel functions $\{J_0(\zeta_i x)\}^{i=1\dots 
\infty}$ (where $\zeta_i$ are the zeros
of $J_0$). These Bessel transforms are defined as
\begin{equation}
     f(r) = \sum_{i=1}^{\infty} f_i \,
     J_{0} \left( \zeta_{i} \frac{r}{R} \right) \;\; ,
\end{equation}
with
\begin{equation}
     f_i = \frac{2}{J_1^2(\zeta_i)}
     \int^1_0 \rho f(\rho R) J_0(\zeta_i \rho) d\rho\;\;.
\end{equation}

Using the Fourier-Bessel coefficients $N^{\bar{p}}_{i}(z,E)$ 
and $q_{i}^{\bar{p}}(z,E)$, there is no conceptual difficulty
to extract solutions of Eq.~(\ref{transport_CR}). We do not
repeat the derivation that can be a bit lenghty. Solutions
for primaries can be found in Ref. \cite{pbar_pbh}, whereas 
the one for secondaries has been discussed in Ref. \cite{PaperII}.

\subsection{Energy losses, tertiary component}
\label{sec:energy_redist}
Following the procedure described e.g. in Ref. \cite{PaperI},
energy losses and diffusive reacceleration lead to a 
differential equation on $N_i(z=0,E)$
\begin{eqnarray}
           \lefteqn{A_iN_i(0,E)={\cal Q}_i(E)} \\
           &&-2h\frac{\partial}{\partial E}
           \left\{b_{loss}(E)N_i(0,E)- K_{EE}\frac{\partial 
N_i(0,E)}{\partial E}
           \right\}\;. \nonumber
           \label{eq_a_resoudre2}
\end{eqnarray}
where $b_{loss} = b_{ion} + b_{Coul} + b_{adiab}$ includes the three 
kinds of energy losses.
The exact forms for these terms may be found in Refs. \cite{PaperI,revue}.
The resolution of this equation proceeds as described in App.~(A.2),
(A.3) and App. (B.) of Ref. \cite{PaperII}, to which we refer for further
details. The source term also takes into account the so-called 
tertiary component,
$q_{i}^{ter}(E)$, corresponding to inelastic but non-annihilating 
reaction of $\bar{p}$
on interstellar matter. This mechanism merely redistributes antiprotons
towards lower energies and tends to flatten their spectrum.

\

{\bf Acknowledgments}. We warmly thank Prof. A. Bottino for very useful
discussions.  N.F. wishes to thank, for the warm hospitality and
support, the Korean Institute for Advanced Study (KIAS), where part of
this work has been done.


\end{document}